\pdfoutput=1
\documentclass[
  journal=jpccck,
  manuscript=article,
  layout=twocolumn,
  email=true,
  ]{achemso}                      % Define documentclass

\usepackage[T1]{fontenc}          % Sets output encoding (fontenc should be called before inputenc)
\usepackage[utf8]{inputenc}       % Sets input encoding (latin1 and utf8 enable umlauts and accents. utf8 is recommended for all operating systems)
\usepackage{csquotes}             % Ensure that quoted texts are typeset according to the rules of your main language
\usepackage{chemmacros}           % Support for typesetting chemical formulae (e.g. IUPAC names with \iupac{name})
	\chemsetup{greek=textgreek}     % Set greek module to use for non-italic greek letters (if not set, chemmacros will give a warning)
\usepackage{siunitx}              % Enables the setting of SI units by using \SI{value}{unit} or \si{unit}
\usepackage{graphicx}             % Enables the embedding of grahpics by using \includegraphics[]{}
\usepackage{float}                % Positioning of images and tables (force postion with [H])
\usepackage{setspace}             % Set space between lines
\usepackage{geometry}
\usepackage{bibunits}             % Multiple bibliographies in one document
  \defaultbibliographystyle{achemso}
\usepackage{hyperref}             % Automatically create hyperlinks (usally, this package should be included as the last package)
  \hypersetup{
    breaklinks=true,              % Allows links to be broken into multiple lines
    colorlinks=true,              % Highlight links with colors instead of boxes
    urlcolor=blue,
  }
  \AtBeginDocument{} % Workaround to get rid of errors when using hyperref and labelling an align environment

\mciteErrorOnUnknownfalse  % Git rid of "Package mciteplus Error: Entry Status for cite key <...> under tracking ID `main' is undefined, treating it as a head entry."

\newcommand{\Lip}{Li$^+$}
\newcommand{\TFSI}{NTf$_2$}
\newcommand{\TFSIm}{[NTf$_2$]$^-$}
\newcommand{\LiTFSI}{Li[NTf$_2$]}
\newcommand{\OBT}{O$_{\text{NTf}_2}$}
\newcommand{\NBT}{N$_{\text{NTf}_2}$}
\newcommand{\OPEO}{O$_\text{PEO}$}

\title{Impact of Charged Surfaces on the Structure and Dynamics of Polymer Electrolytes: Insights from Atomistic Simulations}
\author{Andreas Thum}
\affiliation{Universität Münster, Institut für Physikalische Chemie, Corrensstraße 28/30, 48149 Münster, Germany}
\email{a.thum@uni-muenster.de}
\author{Diddo Diddens}
\affiliation{Helmholtz-Institut Münster: Ionenleitung in der Energiespeicherung (IEK-12), Forschungszentrum Jülich GmbH, Corrensstraße 46, 48149 Münster, Germany}
\email{d.diddens@fz-juelich.de}
\author{Andreas Heuer}
\affiliation{Universität Münster, Institut für Physikalische Chemie, Corrensstraße 28/30, 48149 Münster, Germany}
\alsoaffiliation{Helmholtz-Institut Münster: Ionenleitung in der Energiespeicherung (IEK-12), Forschungszentrum Jülich GmbH, Corrensstraße 46, 48149 Münster, Germany}
\email{andheuer@uni-muenster.de}
\date{Dated: \today}

\begin{document}

\begin{bibunit}
\begin{abstract}
  Polymer electrolytes are intensely investigated for use as solid electrolytes in next generation lithium-ion and lithium-metal batteries. However, little is known about the structural and dynamical properties of polymer electrolytes close to electrode surfaces. Here, a PEO-\LiTFSI{} polymer electrolyte, confined between two oppositely charged graphite-like electrodes, is studied via molecular dynamics simulations. Three different surface charges of $\sigma_S = 0$, $\pm 0.5$ and $\pm \SI{1}{\elementarycharge\per\square\nano\meter}$ are considered. Upon charging, a very strong and component-specific layering is observed. Only for the highest surface charge, lithium ions get desolvated and come into direct contact with the negative electrode. The layer structure goes along with the emergence of free energy barriers, which lead to a reduction of the lithium-ion dynamics, as quantified by spatially resolved mean-square displacements, corrected for a drift component. Interchain transfers that are known to be very important for long-range lithium-ion transport in polymer electrolytes play no significant role for transitions of lithium ions between different layers.
\end{abstract}

\section*{Introduction}

Batteries are ubiquitous in everyday life. The commercialization of lithium-ion batteries in 1991 has lead to an explosion of portable electronic devices, ranging from smartphones and laptops to battery-driven tools and household aids. Other emerging fields for battery technologies are electromobility and the storage of electricity produced by renewable energy sources \cite{Xie2020, Goodenough2013, Whittingham2012, Yoshino2012}.

The next generation of lithium-based batteries is expected to be driven by solid electrolytes, due to important advantages over conventional liquid organic electrolytes. Particularly, solid electrolytes are usually less or even non-flammable and prevent battery leakage \cite{Kalhoff2015}. Moreover, solid electrolytes could possibly suppress the growth of lithium dendrites and therefore enable rechargeable lithium-metal batteries which allow for much higher energy and power densities than lithium-ion batteries \cite{Janek2016, Sun2017}.

Besides inorganic ceramics, organic polymers are promising candidates for solid or quasi-solid electrolytes for lithium-ion and lithium-metal batteries \cite{Sun2017, Manthiram2017}. After the first report of ion conduction in polymers in 1973 \cite{Fenton1973, Wright1975}, poly(ethylene oxide) (PEO) was soon proposed as solid electrolyte for electrochemical applications \cite{Armand1978}. However, neat, linear PEO suffers from low ionic conductivity, which is typically well below $\SI{1e-4}{\siemens\per\centi\meter}$ at room temperature and thus considerably below the minimum of $\SI{1e-3}{\siemens\per\centi\meter}$ required for practical application in lithium-ion batteries. Therefore, a variety of other polymer hosts and architectures was investigated \cite{Long2016, Bresser2019, Mindemark2018}. But despite the drawbacks of high molecular weight PEO, polyethers remain a frequently used building block in new polymer electrolytes \cite{Xue2015, Doerr2018, Pelz2019, Nair2019a, Nematdoust2020, Butzelaar2021}. Moreover, due to its simplicity and in-depth exploration, PEO is an important model and reference system.

Because batteries rely on the electrochemistry taking place at the electrode-electrolyte interface, understanding the impact of this interface on the electrolyte and especially on the lithium-ion transport is crucial for the development of improved battery systems. It is well known that particularly during the first charge-discharge cycle conventional liquid organic electrolytes partially decompose at the electrode surface, because they are thermodynamically unstable against the high-voltage electrodes. The decomposition products form an electrically insulating but ion conducting solid passivation layer on the electrodes, the so-called solid electrolyte interphase (SEI), which kinetically prevents further electrolyte degradation \cite{Peled1979, Peled1997}. Although much progress has been made in elucidating the SEI formed by conventional liquid organic electrolytes, a fundamental understanding how it can be controlled and optimized is still lacking \cite{Winter2009, Xu2011, Peled2017}. Even less is known about the electrode-electrolyte interphase in batteries utilizing solid electrolytes, despite the huge progress made in recent years \cite{Janek2016, Sun2017, Yu2018, Sangeland2019}.

Apart from the composition and morphology of and the lithium-ion transport inside the SEI itself, it is also important to know how the structure of and the lithium-ion transport inside the electrolyte change in the vicinity of the electrode (or SEI) surface. On the one hand, this elucidates which electrolyte components predominate at the surface and thus can decompose and build the SEI. On the other hand, understanding how lithium ions are transported to the surface is essential to improve the performance of the battery.

Molecular dynamics (MD) simulations have become a valuable tool to study the effect of (charged) surfaces on the atomistic structure and dynamics of conventional liquid electrolytes \cite{Wang2018, LyndenBell2007} and ionic liquids (ILs) \cite{LyndenBell2007, Merlet2013a, Fedorov2014}. MD simulations have also been very successful in unraveling the \Lip{}-transport mechanisms in polymer electrolytes \cite{Borodin2006, Maitra2007, Diddens2010} and they have already been employed to study the behavior of neat polymers (without dissolved salts) adsorbed at surfaces \cite{Baschnagel2003, Binder2012}. However, only a very limited number of atomistic simulations addressed the influence of surfaces on the structure and dynamics of polymer electrolytes (with dissolved salts). First simplistic MD simulations of polymer electrolytes near surfaces were performed by Aabloo and Thomas, who simulated three LiCl ion pairs in a PEO slab with tethered chain ends \cite{Aabloo2001}. They found that the Li-anion coordination number decreases when approaching the surface region. Ebadi \latin{et al.} \cite{Ebadi2017} studied a PEO-\LiTFSI{} electrolyte at uncharged lithium-metal surfaces. In their simulation they have seen a more ordered structure of the electrolyte near the surface which was accompanied by a decrease of the overall dynamics. More recently, Louren\c{c}o \latin{et al.} \cite{Lourenco2020} simulated ternary polymer electrolytes composed of PEO, a lithium salt and an ionic liquid confined between two lithium-metal electrodes. They found that the shape of the IL ions significantly influence the layer structure near the electrode surface. The same authors were also involved in two studies using first principle calculations to investigate the initial process of SEI formation between lithium-metal anodes and various polymer electrolytes \cite{Ebadi2019, Mirsakiyeva2019}. However, none of these studies considered charged surfaces and a deeper understanding of the lithium-ion dynamics in polymer electrolytes near (charged) surfaces is still lacking.

To study the effect of uncharged and charged surfaces on the structure of and the lithium-ion transport in polymer electrolytes, we have performed MD simulations of a PEO-\LiTFSI{} polymer electrolyte confined between two model electrodes. The model electrodes are hexagonal lattices of Lennard-Jones spheres, mimicking the basal plane of graphite. The actual chemical nature of the electrodes is not of primary importance for our study, because we are only interested in the electrolyte structure and lithium-ion transport near and on a (charged) surface. To also capture (electro)chemical reactions at interfaces, first-principles calculations or reactive MD simulations would be required \cite{Ebadi2019, Mirsakiyeva2019, Takenaka2014, Biedermann2021a}.

This paper presents our findings in the following order. After a description of the simulation setup, we discuss the influence of uncharged and charged surfaces on the structural properties of the electrolyte, including the formation of layers, the \Lip{}-coordination environment and the internal structure of the layers. Subsequently, we discuss the influence of the layer structure on the \Lip{}-transport perpendicular and parallel to the electrode surfaces. Thereby we also evaluate the role of interchain transfers for transitions of lithium ions between different layers.

\section*{Simulation Details}

All simulations have been performed with the Gromacs 2018 software package \cite{Berendsen1995_GROMACS, VanDerSpoel2005_GROMACS, Abraham2015_GROMACS, Abraham2019_GROMACS} using the OPLS-AA force field \cite{Jorgensen1996_OPLS-AA} for PEO and the OPLS-AA-derived CL\&P force field\cite{Gouveia2017_CLP, CanongiaLopes2012_CLP} for \LiTFSI{}. Atomic point charges of ionic species were scaled by a factor of $0.8$ to account for polarization effects in a mean-field like manner. This charge scaling approach, whose theoretical foundation was given by Leontyev and Stuchebrukhov \cite{Leontyev2011}, has become quite popular in the last years for systems with high ion densities, because it enables an effective description of polarization effects without additional computational cost \cite{Dommert2012, Schroeder2012, Salanne2015}. We have validated the charge scaling approach for our particular PEO-\LiTFSI{} system by comparing different charge scaling schemes to simulations using the APPLE\&P polarizable force field \cite{Borodin2006_APPLEP, Borodin2006_APPLEPa, Borodin2009_APPLEP} and to experiments (see the section \enquote{Force field validation} in the supporting information). We did not employ the APPLE\&P polarizable force field for our simulations, because our systems are relatively big and the incorporation of explicit polarization is computationally expensive.

\begin{figure}[!ht]
  \centering
  \includegraphics[width=\columnwidth]{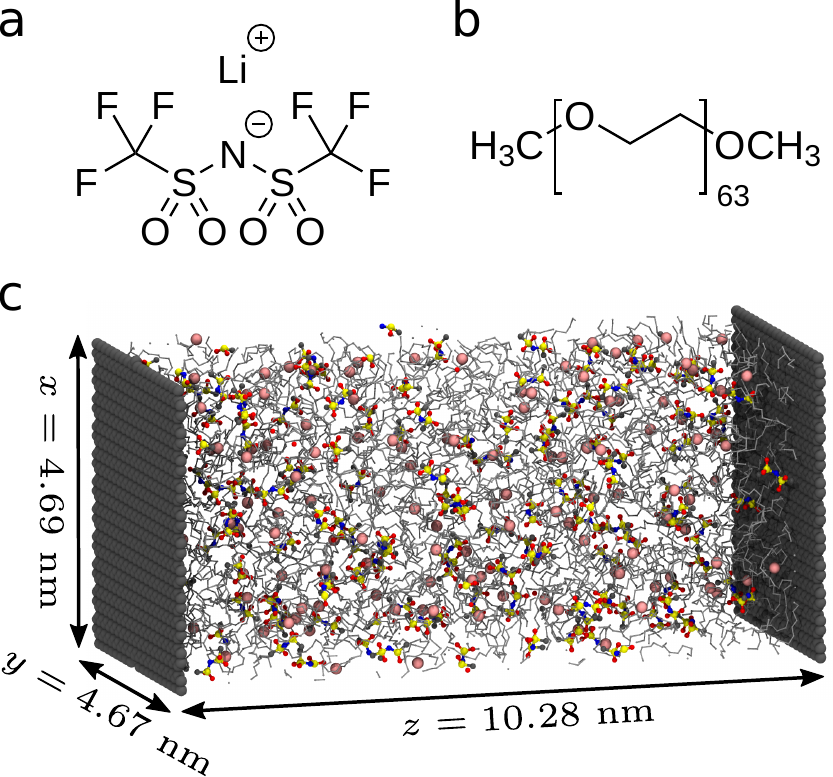}
  \caption{Structures of (a) lithium \iupac{bis(trifluoromethanesulfonyl)imide} (\LiTFSI{}) and (b) \iupac{poly(ethylene oxide) dimethyl ether} (in this paper abbreviated as PEO; another common abbreviation is PEGDME for \iupac{poly(ethylene glycol) dimethyl ether}). (c) Snapshot of the simulation box with box dimensions. Carbon atoms are gray, nitrogen atoms blue, oxygen atoms red,  sulfur atoms yellow and lithium ions pink. PEO chains are sketched as silver lines. Hydrogen and fluorine atoms are omitted.}
  \label{fig:simulation_setup}
\end{figure}

To investigate the influence of uncharged and charged surfaces on a PEO-\LiTFSI{} polymer electrolyte, the following simulation protocol was developed. $128$ \LiTFSI{} ion pairs and $40$ methoxy-terminated, coiled PEO chains with $n = 63$ monomers (i.e.\ $64$ ether oxygens) each were randomly packed into a box using the packmol software \cite{Martinez2009}. This yields an ether oxygen to lithium ratio of EO:Li = 20:1 and a molecular weight of the PEO chains of $M_w = \SI{2821.43}{\gram\per\mole}$. The structures of \LiTFSI{} and methoxy-terminated PEO are shown in Figure~\ref{fig:simulation_setup}a and b, respectively. The $x$- and $y$-dimensions of the box were already adjusted to the size of the graphene sheets, which will be inserted later, and were not allowed to change during the whole simulation procedure. Also the extra space needed to ensure a proper periodic continuation of the graphene sheets was already incorporated.

The such generated starting configuration underwent an energy minimization, followed by \SI{3}{\nano\second} of equilibration in an $NVT$ ensemble at \SI{273}{\kelvin} using Langevin dynamics with a high friction coefficient of $f_a = \frac{m_a}{\SI{0.01}{\pico\second}}$ ($a$ stands for the atom type) to further relax unfavorable structures. Periodic boundary conditions were only applied in $x$- and $y$-direction, whereas in $z$-direction the simulation box was limited by two implicit walls. The walls interact with the polymer electrolyte via a 10--4 Lennard-Jones potential, which mimics a uniform, two-dimensional density of Lennard-Jones particles on the wall surface. The Lennard-Jones parameters of the walls were taken from the OPLS-AA atom-type opls147 ($\sigma = \SI{0.355}{\nano\meter}$, $\epsilon = \SI{0.29288}{\kilo\joule\per\mole}$) that was parametrized for fused benzene rings. This atom type was also used for the graphene sheets, which are inserted in a later step. The wall surface density was set to $\SI{38.177}{\per\square\nano\meter}$, which is the surface density of atoms in a graphene lattice. For Coulomb interactions a plain cutoff at $\SI{1.4}{\nano\meter}$ was used, whereas van-der-Waals forces were smoothly shifted to zero between 1.3 and $\SI{1.4}{\nano\meter}$ and long-range dispersion corrections for energy and pressure were applied \cite{Abraham2019_GROMACS}. A stochastic leap-frog integrator \cite{Goga2012} was employed for integrating Langevin's equation with a time step of $\SI{0.5}{\femto\second}$. Hydrogen bonds were kept constrained by the linear constraint solver (LINCS) algorithm \cite{Hess1997} with a fourth-order expansion of the constraint coupling matrix and two iterations per time step. A possibly accumulated translational velocity of the center of mass of the system was removed every $\SI{0.2}{\pico\second}$. Another equilibration step of $\SI{3}{\nano\second}$ in an $NVT$ ensemble followed, but this time with a usual leap-frog integrator \cite{Abraham2019_GROMACS} for Newtonian dynamics and a velocity-rescale thermostat \cite{Bussi2007}. The reference temperature was raised to $\SI{423}{\kelvin}$ with a coupling time of $\SI{0.5}{\pico\second}$. Coulomb interactions beyond the cutoff were calculated by the smooth particle mesh Ewald (SPME) method \cite{Darden1993, Essmann1995} with the Yeh and Berkowitz correction for slab geometries \cite{Yeh1999}, using an implicit scaling factor of three for the box height. The grid spacing for the fast Fourier transformation was set to $\SI{0.12}{\nano\meter}$, the interpolation order to six and the relative strength of the electrostatic interactions at the Coulomb cutoff to $1 \times 10^{-5}$. In a next step, Berendsen pressure scaling \cite{Berendsen1984} was applied for $\SI{3}{\nano\second}$ along the $z$ axis to shrink the box and obtain a correct density. The reference pressure was set to $\SI{1}{\bar}$, the coupling constant to $\SI{0.5}{\pico\second}$ and the compressibility to $\SI{4.5e-5}{\per\bar}$. To ensure a proper mixing of the polymer electrolyte, the dynamics were accelerated by elevating the temperature to $\SI{623}{\kelvin}$ for $\SI{100}{\nano\second}$. The time step was increased to $\SI{2}{\femto\second}$ and the coupling times of the velocity-rescale thermostat and the Berendsen barostat were both increased to $\SI{1}{\pico\second}$. The number of LINCS iterations per time step was decreased to one. Subsequently, the system was allowed to relax again for $\SI{100}{\nano\second}$ at $\SI{423}{\kelvin}$.

After this initial equilibration procedure, the actual graphene surfaces were inserted, which have been build using VMD's \cite{Humphrey1996_VMD} carbon nanostructure builder plugin. The C--C bond length was set to $r_0 = \SI{0.142}{\nano\meter}$, the edge type to zigzag and the size to $4.6 \times \SI{4.6}{\nano\meter}$. This has lead to an actual size of the graphene sheets of $4.54 \times \SI{4.55}{\nano\meter}$. Taking into account that the sheets are periodically continued, additional space of $r_0$ in $x$- and $r_0 \cdot \sin(\SI{60}{\degree})$ in $y$-direction has to be added, leading to $xy$-dimensions of the simulation box of $4.69 \times \SI{4.67}{\nano\meter}$. Because the carbon atoms of the graphene sheets were fixed, also the $z$-dimension of the box cannot change anymore after insertion of the graphene sheets (i.e.\ pressure scaling is not possible anymore), making it necessary to adjust the $z$-direction beforehand to ensure a correct average density. Therefore (and for reference reasons), also a bulk simulation of the system was performed using basically the same procedure and settings as above, but utilizing a cubic simulation cell with periodic boundary conditions in all three dimensions, normal SPME electrostatics and isotropic pressure scaling. From the last $\SI{10}{\nano\second}$ of the last equilibration step of this bulk simulation, the density was calculated to be $\SI{1143.36}{\kilogram\per\cubic\meter}$. Given the fixed $x$- and $y$-dimensions of the system with graphene sheets and regarding the Lennard-Jones radius $\sigma$ of the graphene carbon atoms as excluded volume, the $z$-dimension of the box had to be set to $\SI{10.28}{\nano\meter}$ to resemble the same total density as in the bulk simulation. Figure~\ref{fig:simulation_setup}c shows a snapshot of the final simulation box.

After insertion of the graphene sheets, the implicit walls were turned off and another energy minimization was performed. In a last step, the system was simulated for $\SI{1.1}{\micro\second}$ using a Nosé-Hoover thermostat \cite{Nose1984, Hoover1985} for the polymer electrolyte (the graphene sheets were kept fixed and were thus not coupled to the thermostat). The trajectory was written out every $\SI{2}{\pico\second}$. The first $\SI{100}{\nano\second}$ were discarded as further equilibration and the last $\SI{1}{\micro\second}$ was used as production run.

A snapshot after $\SI{500}{\nano\second}$ of the production run was used to start simulations with charged surfaces. Charges were applied to the surface by uniformly giving each atom of the graphene sheet the same fixed charge. Two simulations with charged surfaces were performed, one with a charge of $\pm \SI{0.0131}{\elementarycharge}$ per graphene atom and one with $\pm \SI{0.0262}{\elementarycharge}$. Given the surface number density of graphene of $\SI{38.177}{\per\square\nano\meter}$, this corresponds to surface charge densities of $\sigma_S = \pm \SI{0.5}{\elementarycharge\per\square\nano\meter}$ and $\sigma_S = \pm \SI{1}{\elementarycharge\per\square\nano\meter}$. The surface at $z = \SI{0}{\nano\meter}$ always got the positive charge and the surface at $z = \SI{10.28}{\nano\meter}$ the negative one.

The cubic simulation box of the bulk system was adjusted to reproduce the average total bulk density, too, and an $NVT$ simulation of $\SI{1.1}{\micro\second}$ length with basically the same settings as used for the surface simulations was performed.

All analyses were done using Gromacs tools \cite{Abraham2019_GROMACS} and the MDAnalysis Python package \cite{Michaud-Agrawal2011_MDAnalysis, Gowers2016_MDAnalysis}. Plots were created with the Matplotlib Python package\cite{Hunter_2007_matplotlib} and the snapshots were rendered with VMD \cite{Humphrey1996_VMD, Stone1998_MAthesis_VMD}.

\section*{Results and Discussion}
\subsection*{Electrolyte Structure}
\subsubsection*{Layering}

\begin{figure*}[!ht]
  \centering
  \includegraphics[width=\textwidth]{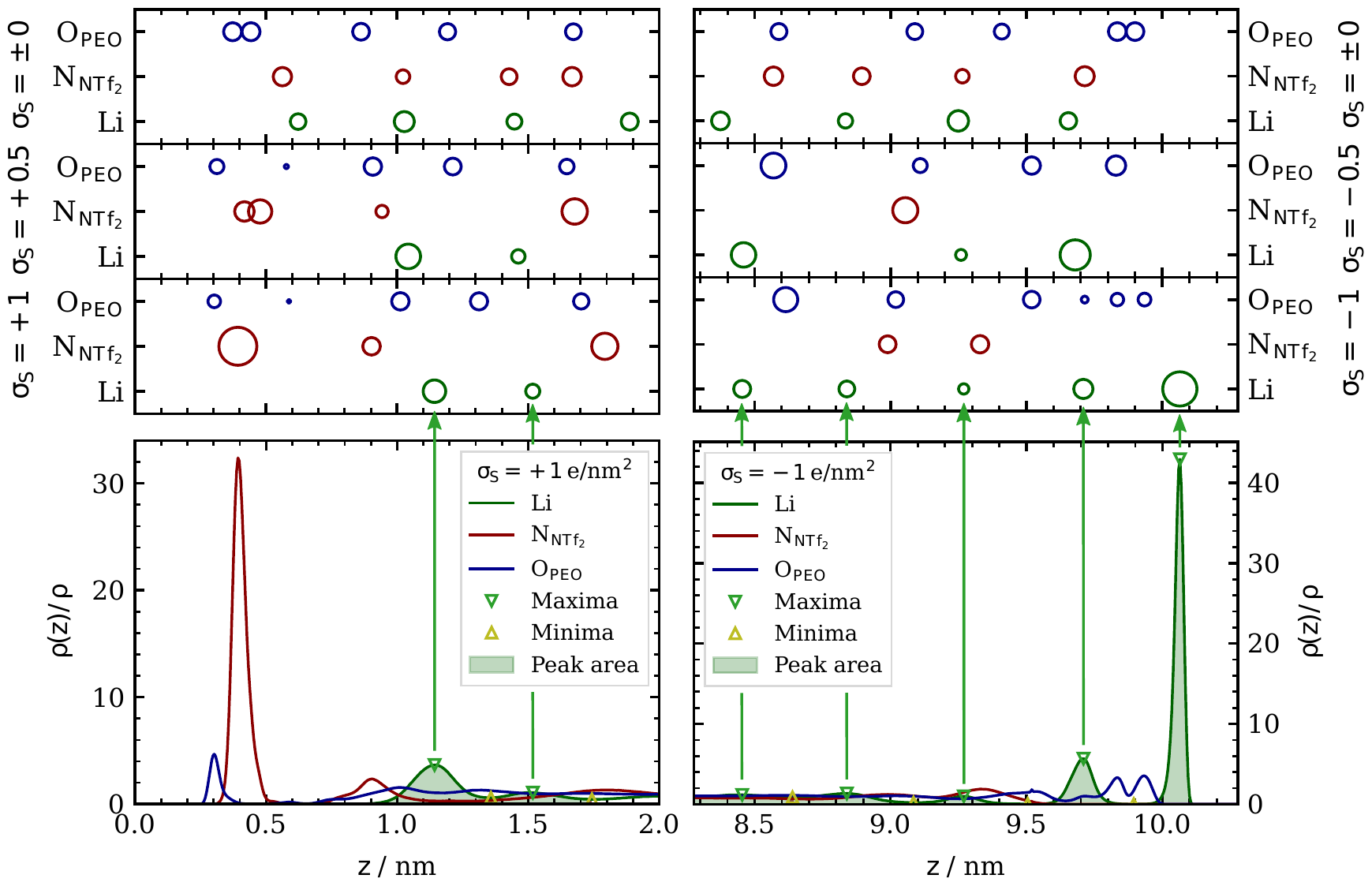}
  \caption{Top panel: Peak positions of the density profiles (see Figures~\ref{supp:fig:dens-z} in the SI) for different electrolyte components and for the three investigated surface charges $\sigma_S = 0$, $\pm 0.5$ and $\pm \SI{1}{\elementarycharge\per\square\nano\meter}$. The area of the circles is proportional to the area below the respective peak and therefore indicative for the number density of the corresponding species at this point relative to its bulk density. Bottom panel: Sketch showing how the upper plot was obtained from the density profiles. The positive electrode resides at $z = \SI{0}{\nano\meter}$ and the negative one at $z = \SI{10.28}{\nano\meter}$, which is the right boundary of the right plots.}
  \label{fig:dens-z_peak_positions}
\end{figure*}
The graphene surfaces induce a layering of the electrolyte components in their immediate vicinity due to the unlike interactions between the different electrolyte components with the surface and with each other. The layering also occurs at uncharged surfaces, but changes dramatically with increasing surface charge. This can be seen in the upper panel of Figure~\ref{fig:dens-z_peak_positions}, which illustrates the peak positions and intensities of the density profiles of \Lip{}, the nitrogen atoms of \TFSI{} (\NBT{}) and of the oxygen atoms of PEO (\OPEO{}) perpendicular to the surfaces (i.e.\ in $z$-direction) for the three investigated surface charges $\sigma_S = 0$, $\pm 0.5$ and $\pm \SI{1}{\elementarycharge\per\square\nano\meter}$. The original density profiles are shown in Figure~\ref{supp:fig:dens-z} in the supporting information (SI). The scatter plot was rendered by first identifying the positions of the maxima and minima in the density profiles based on a minimum threshold for their prominences and widths using SciPy routines \cite{Virtanen2020_SciPy}, namely the \textit{find\_peaks} function of SciPy's \textit{signal} module. Afterwards, the peaks were integrated between the two next nearest minima to get a measure for the number density of the respective electrolyte component in the considered layer relative to its bulk density. This procedure is sketched in the lower panel of Figure~\ref{fig:dens-z_peak_positions}. Finally, the scatter plot was drawn by assigning one circle to each maximum with the circle area proportional to the peak area.

\begin{figure*}[!ht]
  \centering
  \includegraphics[width=\textwidth]{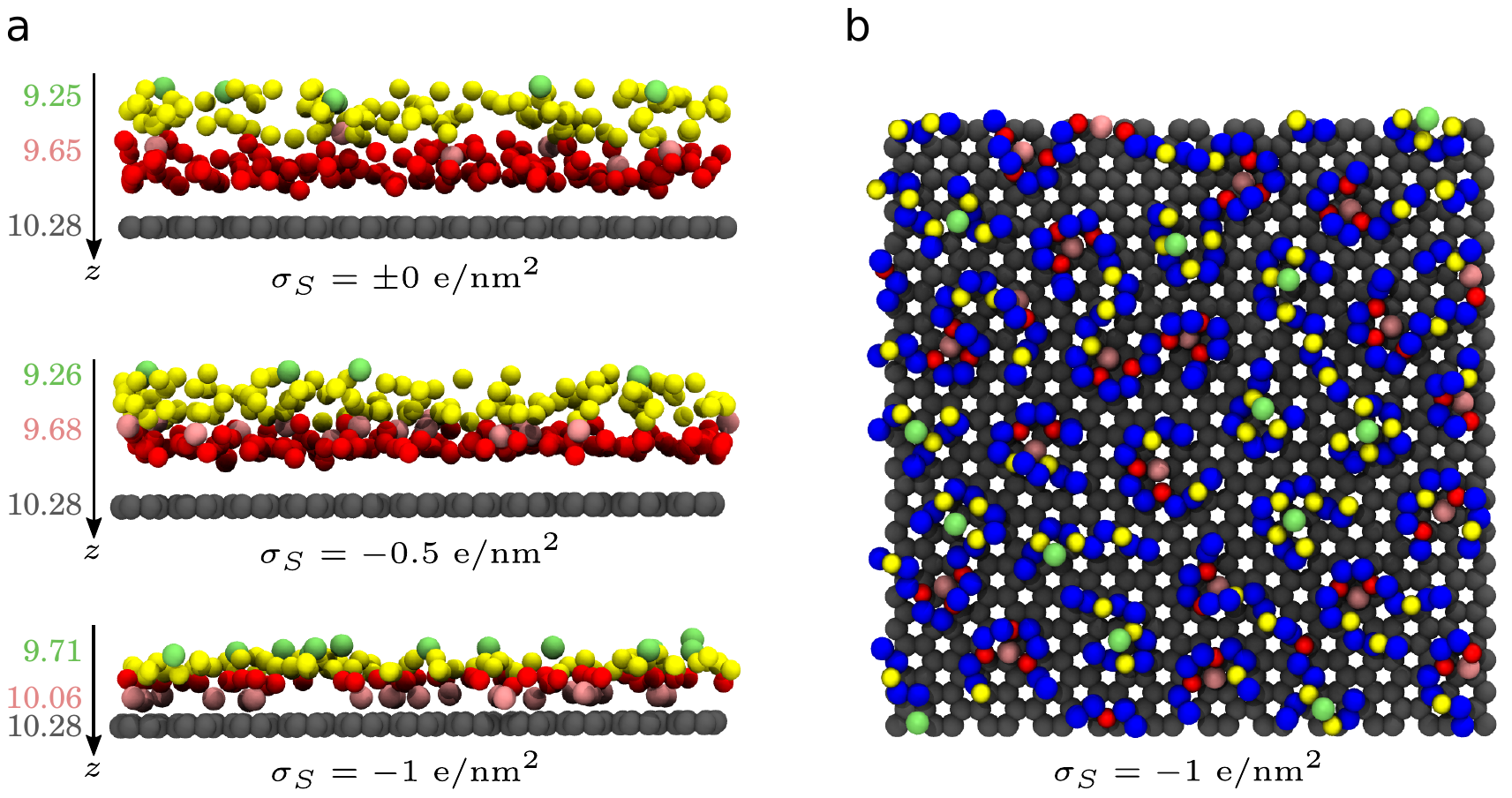}
  \caption{Snapshots of the first two \Lip{}- and \OPEO{}-layers at the negative electrode. (a) Side view for the three investigated surface charges. The $z$-values to the left indicate the positions of the peak maxima of the first two \Lip{}-layers and the position of the negative electrode. (b) Top view for $\sigma_S = -\SI{1}{\elementarycharge\per\square\nano\meter}$. Lithium ions and PEO oxygens in the first layer are colored pink and red, respectively, lithium ions and PEO oxygens in the second layer are colored green and yellow. In (b), carbon atoms of PEO are colored blue.}
  \label{fig:snaphot_positive_electrode}
\end{figure*}

As can be seen in the upper panel of Figure~\ref{fig:dens-z_peak_positions}, at neutral graphene surfaces the first \Lip{}-layer only appears after an \OPEO{}- and a \NBT{}-layer. This behavior was also observed for [Li(G4)][\TFSI{}] solvate ionic liquids near graphene surfaces (G4 = tetraglyme) \cite{Coles2017} and for PEO-\LiTFSI{} electrolytes with and without ionic liquid additives near uncharged lithium surfaces \cite{Ebadi2017,Lourenco2020}. With increasing surface charge, counterions are attracted and coions are repelled by the electrodes. Additionally, the first layers get densified, which can be seen from the increasing circle areas of the first counterion-layers and a shift of the counterion- and \OPEO{}-layers towards the electrodes. However, even at $\sigma_S = -\SI{0.5}{\elementarycharge\per\square\nano\meter}$ the lithium ions are not able to directly approach the electrode, again in agreement with observations by Coles \latin{et al.} \cite{Coles2017} for [Li(G4)][\TFSI{}] solvate ionic liquids. Apparently, the electrostatic attraction of electrodes with $\sigma_S \leq -\SI{0.5}{\elementarycharge\per\square\nano\meter}$ is too weak to desolvate lithium ions from chelating ether ligands. Only at a surface charge of $\sigma_S = -\SI{1}{\elementarycharge\per\square\nano\meter}$, lithium ions get desolvated and directly approach the electrode. This is also nicely demonstrated by the snapshots of the first two \Lip{}- and \OPEO{}-layers shown in Figure~\ref{fig:snaphot_positive_electrode}a. Additionally, Figure~\ref{fig:snaphot_positive_electrode}a illustrates the densification of the first layers with increasing surface charge.

The emerging \Lip{}-layer directly attached to the negative electrode with $\sigma_S = -\SI{1}{\elementarycharge\per\square\nano\meter}$ leads to a splitting of the \OPEO{}-layer, which was formerly located at $z = \SI{9.83}{\nano\meter}$ for $\sigma_S = -\SI{0.5}{\elementarycharge\per\square\nano\meter}$, into a double layer and a small shoulder, as can be seen in the upper and lower panel of Figure~\ref{fig:dens-z_peak_positions}. The PEO oxygens in the first branch of this double layer coordinate to the lithium ions directly attached to the electrode, whereas the oxygen atoms in the second branch coordinate mainly to the lithium ions in the second \Lip{}-layer, as indicated by Figure~\ref{fig:snaphot_positive_electrode}a and b. Additionally, Figure~\ref{fig:snaphot_positive_electrode}b demonstrates that the PEO chains wrap around the lithium ions on the surface and in the second \Lip{}-layer, resulting in a locally anisotropic structure of the polymer chains.

At high surface charges of $\sigma_S = \pm \SI{1}{\elementarycharge\per\square\nano\meter}$ two strongly populated counterion-layers are necessary to screen the electrodes. From Figure~\ref{fig:dens-z_peak_positions} it seems that this is the case for $\sigma_S = \pm\SI{0.5}{\elementarycharge\per\square\nano\meter}$, too. However, a comparison with Figure~\ref{supp:fig:dens-z} in the SI shows that in this case the second counterion-layer is actually buried in the following coion-layer so that effectively there is only one counterion-layer.

Another characteristic that can be seen from the upper panel of Figure~\ref{fig:dens-z_peak_positions} is that a \Lip{}-layer is usually surrounded by two \OPEO{}-layers. In this way, the lithium ions can be coordinated by PEO oxygens from both sides. This sandwich structure can also be seen in the snapshots of Figure~\ref{fig:snaphot_positive_electrode}a.

To sum up, the presence of the surface leads to a complex layering behavior of the electrolyte components which strongly depends on the surface charge and reaches \SIrange{3.0}{3.5}{\nano\meter} into the bulk.

\subsubsection*{Lithium-Ion Coordination}

In the bulk simulation, a lithium ion is on average coordinated by six PEO oxygens from one PEO chain (blue circles in the right panel of Figure~\ref{supp:fig:coordination_histogram_Li-PEO} and bold numbers in Table~\ref{supp:tab:coordination_numbers_Li-PEO} in the SI). Hardly any lithium ion is coordinated by \TFSI{} anions (Figure~\ref{supp:fig:coordination_histogram_Li-NTf2} and Table~\ref{supp:tab:coordination_numbers_Li-NTf2} in the SI). This coordination environment is overall in good agreement with simulations of PEO-\LiTFSI{} electrolytes utilizing the APPLE\&P polarizable force field \cite{Diddens2017}. For a more thorough discussion see the section \enquote{Force field validation} in the SI. For the calculation of coordination numbers, a lithium ion is considered to be coordinated to an oxygen atom from \TFSI{} or PEO, when the distance between the lithium ion and the oxygen atom is less than or equal to $\SI{0.3}{\nano\meter}$. This threshold was taken from the first minimum of the Li-oxygen radial distribution functions (Figure~\ref{supp:fig:rdf} in the SI).

\begin{figure*}[!ht]
  \centering
  \includegraphics[width=\textwidth]{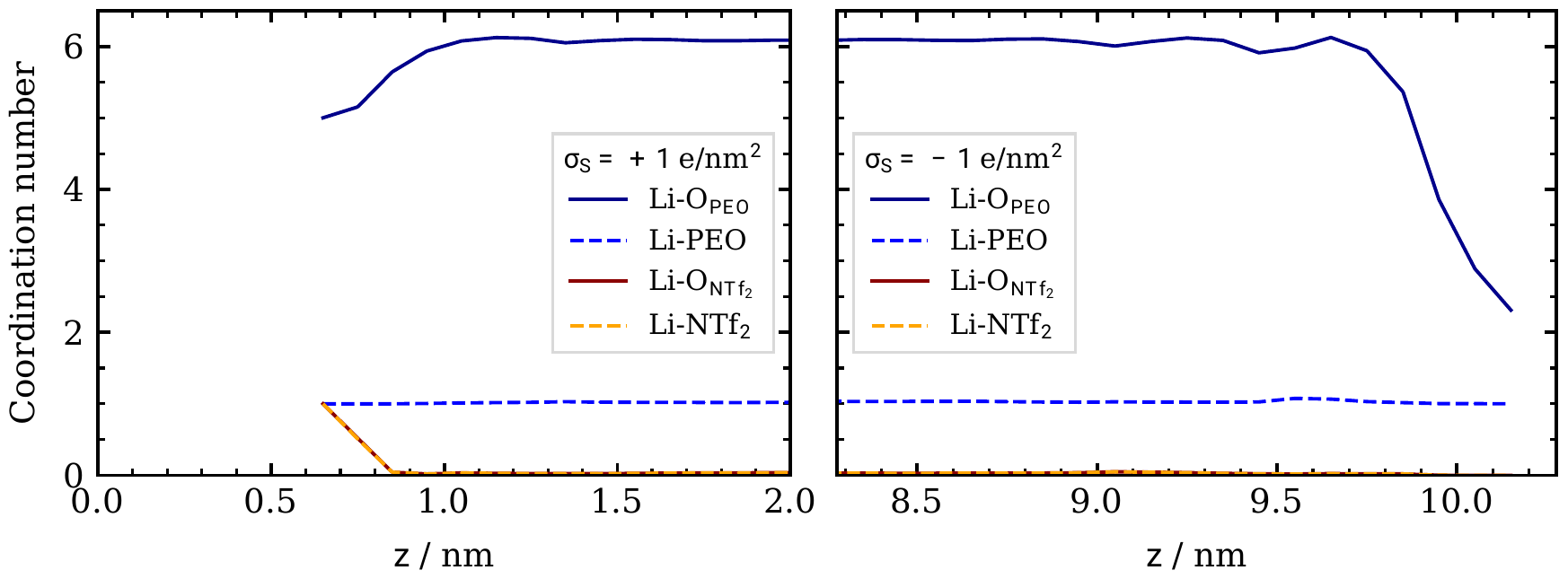}
  \caption{Average lithium-ion coordination numbers as function of $z$ for $\sigma_S = \pm \SI{1}{\elementarycharge\per\square\nano\meter}$. Li-\OBT{} denotes coordinations of \Lip{} to single oxygen atoms from \TFSI{}. Li-\TFSI{} denotes coordinations of \Lip{} to entire \TFSI{} anions. The definitions of Li-\OPEO{} and Li-PEO are analogous. The corresponding plots for the other two surface charges are shown in Figure~\ref{supp:fig:coordination_number_z} in the SI.}
  \label{fig:coordination_number_z}
\end{figure*}
In the bulk region of the surface simulations, lithium ions have the same coordination environment as in the bulk simulation, what can be seen in Figure~\ref{fig:coordination_number_z}. This Figure shows average \Lip{} coordination numbers as function of $z$ for the surface simulation with $\sigma_S = \pm \SI{1}{\elementarycharge\per\square\nano\meter}$. The corresponding plots for the other two surface charges are shown in Figure~\ref{supp:fig:coordination_number_z} in the SI. Figure~\ref{fig:coordination_number_z} demonstrates that in the bulk region a lithium ion is on average coordinated by approximately six oxygens, all of which originate from one PEO chain and almost none from an \TFSI{} anion. In fact, a lithium ion always stays attached to on average one PEO chain, regardless of its $z$-position. This can be seen from the Li-PEO coordination number which is almost uniformly one across the entire simulation box. The overlapping Li-\OBT{} and Li-\TFSI{} coordination numbers show that the Li-anion coordination is almost exclusively monodentate, which was also observed in the bulk simulation (Figure~\ref{supp:fig:coordination_histogram_Li-NTf2} and Table~\ref{supp:tab:coordination_numbers_Li-NTf2}).

Near the negative electrode, the Li-\OPEO{} coordination number decreases sharply from six to $2.5$. This is in agreement with the finding above that at a surface with $\sigma_S = -\SI{1}{\elementarycharge\per\square\nano\meter}$, lithium ions get desolvated and directly approach the surface. The surface seems to cover half of the four to six interaction sites of a lithium ion such that only two to three interaction sites remain for coordination to PEO oxygens.

Near the positive electrode, the change of the Li-oxygen coordination is less dramatic. Here, the lithium ions substitute an oxygen from PEO by an oxygen from \TFSI{}, which is reasonable because of two facts. First, the number density of \TFSI{} anions and therefore the chance to find an \TFSI{} anion is higher at the positive electrode than in the bulk (see Figure~\ref{fig:dens-z_peak_positions}). Second, the positive lithium ions are electrostatically repelled from the positive electrode. In conjunction with an \TFSI{} anion they can form an electrically neutral complex to reduce the repulsion. However, it must be kept in mind that there are only very few lithium ions near the positive electrode. Hence, the calculated coordination number is statistically less significant in this region.

\subsubsection*{Internal Structure of the First Layer ($\sigma_S = -\SI{1}{\elementarycharge\per\square\nano\meter}$)}

\begin{figure*}[!ht]
  \centering
  \includegraphics[width=\textwidth]{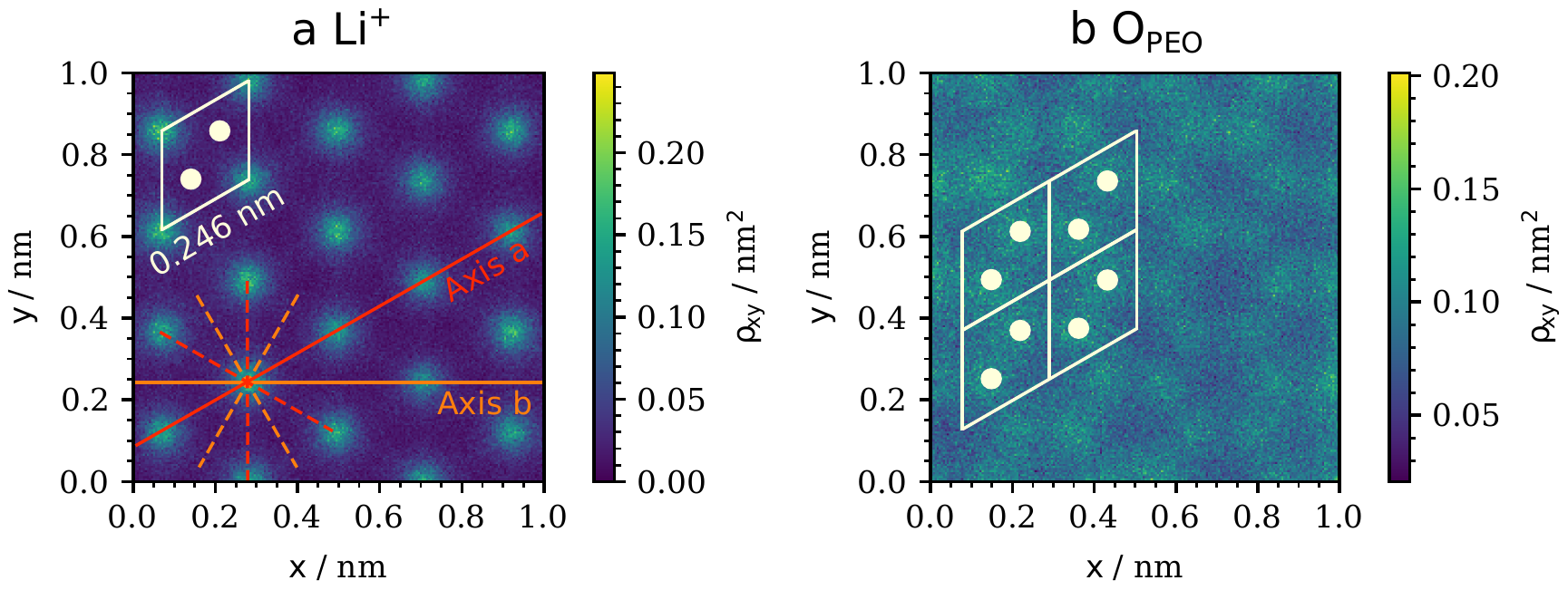}
  \caption{Density maps in $xy$-plane for (a) \Lip{} and (b) \OPEO{} in the first \Lip{}- and the first \OPEO{}-layer (both $z > \SI{9.9}{\nano\meter}$) at the negative electrode with $\sigma_S = -\SI{1}{\elementarycharge\per\square\nano\meter}$. Because the patterns continue periodically, only the first square nanometer of the density maps is shown. In the upper left of the \Lip{}-density map, the unit cell of graphene is sketched in white. Additionally, first-nearest neighbor axes are indicated in red (\enquote{Axis a}) and second-nearest neighbor axes in orange (\enquote{Axis b}). Figure~\ref{fig:free_energy_hex} shows free energy profiles along these kinds of axes. On the left of the \OPEO{}-density map, a graphene supercell is sketched in white.}
  \label{fig:densmap-xy_negative_electrode}
\end{figure*}

In the system with a surface charge of $\sigma_S = \pm\SI{1}{\elementarycharge\per\square\nano\meter}$, the lithium ions which directly approach the negative surface adopt the hexagonal lattice structure of the graphene electrode. Figure~\ref{fig:densmap-xy_negative_electrode}a shows the two-dimensional density distribution of lithium ions in the first \Lip{}-layer ($z > \SI{9.9}{\nano\meter}$) averaged over the entire simulation time. In the upper left of this density map, the unit cell of the graphene electrode is sketched in white. Clearly, the lithium ions favor the centers of the graphene hexagons, probably because there they experience the maximum attraction of six surface atoms.

Figure~\ref{fig:densmap-xy_negative_electrode}b shows the two-dimensional density distribution of PEO oxygens in the first branch of the double layer (also for $z > \SI{9.9}{\nano\meter}$). These oxygen atoms coordinate to the lithium ions in the first \Lip{}-layer, as seen in Figure~\ref{fig:snaphot_positive_electrode}. Their distribution is by far more diffuse, although one can still identify a weak pattern. The PEO oxygens slightly favor to sit directly above the surface atoms rather than within the center of a graphene hexagon.
This can be explained by the system's attempt to minimize the electrostatic repulsion between the negatively charged PEO oxygens and the negative electrode. For a single PEO oxygen, it would be most favorable to sit directly above a lithium ion. But in this way, each lithium ion could be coordinated by only one PEO oxygen. Because there are more PEO oxygens than lithium ions, many PEO oxygens would reside at the negative electrode without being screened by a positive lithium ion. In order to minimize the energy of the entire system, the PEO oxygens must coordinate to the lithium ions diagonally from above, as illustrated in Figure~\ref{fig:snaphot_positive_electrode}b. This allows two to three oxygens to coordinate to the same lithium ion, while still maximizing their distance to the surface atoms.

As can be guessed from the already diffuse \OPEO{}-distribution, a clear internal structure in the following layers was not found. A similar situation is encountered at the positive electrode, where the first \OBT{}- and \OPEO{}-layer are internally structured (Figure~\ref{supp:fig:densmap-xy_positive_electrode} in the SI) but the following layers are not. This means that although the presence of the highly charged electrodes induces the formation of a layer structure reaching several nanometers into the bulk region, the influence of the electrodes on the internal structure of these layers is limited to the first counterion-layer and the first \OPEO{}-layer.

\subsection*{Charge Transport}
\subsubsection*{Free Energy Profiles}
\paragraph*{Perpendicular to the Electrodes.}

\begin{figure*}[!ht]
  \centering
  \includegraphics[width=\textwidth]{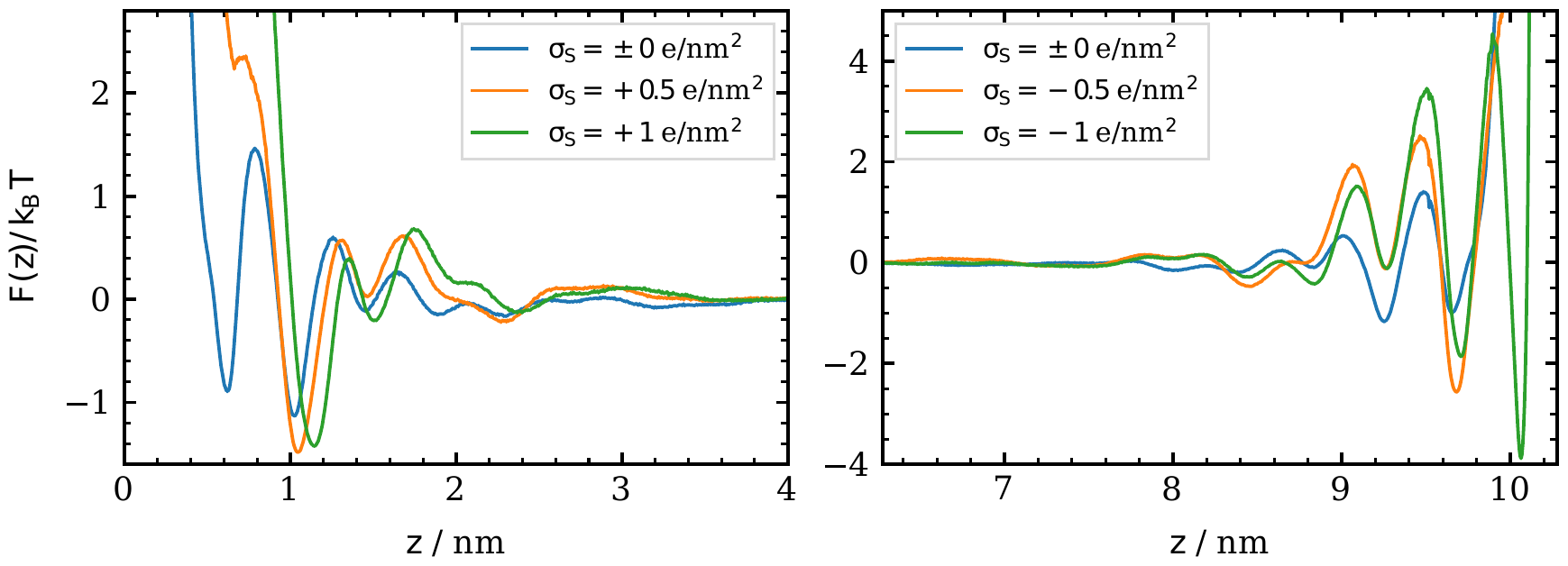}
  \caption{Free energy profiles estimated from the \Lip{}-density profiles (Figure~\ref{supp:fig:dens-z} in the SI) for the three investigated surface charges. The positive electrode resides at $z = \SI{0}{\nano\meter}$ and the negative one at $z = \SI{10.28}{\nano\meter}$, which is the right boundary of the right plot.}
  \label{fig:free_energy_z}
\end{figure*}

The layer structure described above is accompanied by the emergence of free energy barriers. The free energy profile for the lithium-ion transport perpendicular to the electrodes (i.e.\ in $z$-direction) can be estimated from the negative logarithm of the \Lip{}-density profile according to
\begin{equation}
F(z) = -k_B T \ln{[\rho(z)]}.
\label{eq:free_energy}
\end{equation}
Here, $k_B$ is the Boltzmann constant and $T$ is the temperature. The resulting free energy profiles are shown in Figure~\ref{fig:free_energy_z} for the three investigated surface charges. The free energy of lithium ions in the bulk was set to zero as reference point.

In the case of uncharged surfaces, the free energy barrier for leaving the first \Lip{}-layer near the surface is slightly lower than for entering it, because the free energy of lithium ions in the first layer is slightly higher than in the second layer. Thus, a given lithium ion moves more likely out of the first layer than into it. In the case of charged surfaces, the opposite is true: the free energy barrier for leaving the first \Lip{}-layer near either of the two surfaces is higher than for entering it. Hence, at charged surfaces a given lithium ion moves more likely into the first layer than out of it.

Interestingly, the free energy barrier for transitions from the \Lip{}-layer at around \SIrange{1.0}{1.2}{\nano\meter} to the one at \SIrange{1.4}{1.5}{\nano\meter}, i.e.\ the free energy barrier for escaping from the neutral or positive electrode, is very similar in all systems. This barrier arises from the attraction that the lithium ions experience from the first anion layer at the neutral or positive electrode, which seems to be similar in all systems.

At the negative electrode, the free energy barriers for approaching the electrode surface increase significantly with increasing surface charge due to a depletion of lithium ions between the different \Lip{}-layers. Apparently, the layers of the other electrolyte components become less permeable for lithium ions. The barriers for escaping from the negative electrode are even higher, because here the lithium ions must additionally overcome the electrostatic attraction of the electrode. The highest free energy barrier occurs between the first and second \Lip{}-layer at the negative electrode with $\sigma_S = - \SI{1}{\elementarycharge\per\square\nano\meter}$, which can be attributed to the structural reorganization that has to take place when the lithium ions get (partially) desolvated or solvated, depending on whether they approach or escape from the negative surface.

\paragraph*{On the Electrode Surface.}

\begin{figure}[!ht]
  \centering
  \includegraphics[width=\columnwidth]{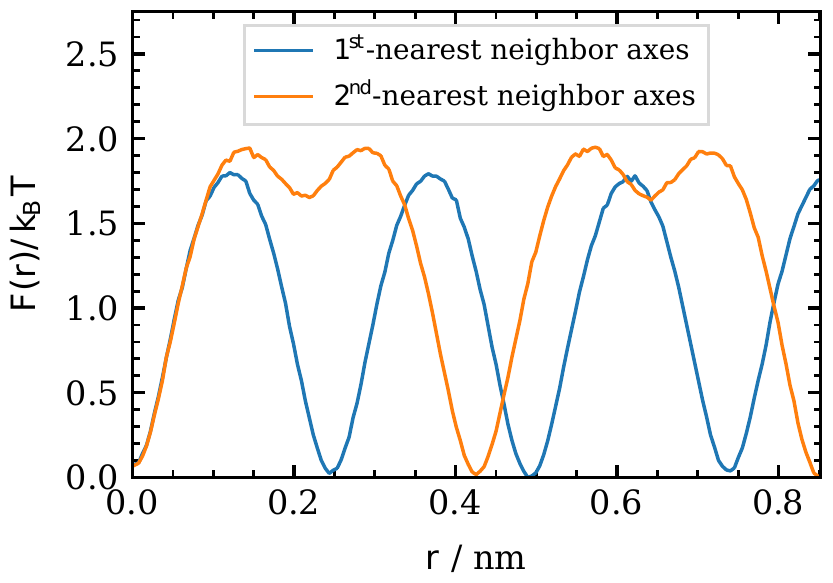}
  \caption{Free energy profiles estimated from the \Lip{}-density profiles (Figure~\ref{supp:fig:density_hex} in the SI) along the first- and second-nearest neighbor axes of the hexagonal lithium sites in the first \Lip{}-layer ($z > \SI{9.9}{\nano\meter}$) at the negative electrode with $\sigma_S = -\SI{1}{\elementarycharge\per\square\nano\meter}$, as indicated in Figure~\ref{fig:densmap-xy_negative_electrode}a. The shown profiles are the average over all first and second-nearest neighbor axes, respectively, and over all trajectory frames. Because the profiles continue periodically, only the first $\SI{0.852}{\nano\meter}$ are shown, which is six times the C-C bond length in graphene ($\SI{0.142}{\nano\meter}$). The origin $r = \SI{0}{\nano\meter}$ is set to the center of a graphene hexagon.}
  \label{fig:free_energy_hex}
\end{figure}

As shown in Figure~\ref{fig:densmap-xy_negative_electrode}a, lithium ions that are in direct contact with the electrode adapt the surface structure of the electrode. This structural ordering indicates the presence of free energy barriers for the diffusion of lithium ions on the electrode surface. These free energy barriers within the first \Lip{}-layer at the negative electrode with $\sigma_S = - \SI{1}{\elementarycharge\per\square\nano\meter}$ can be estimated similar to Equation~\eqref{eq:free_energy} from the density profiles along the hexagonal axes of the distinct lithium sites as indicated in Figure~\ref{fig:densmap-xy_negative_electrode}a. The density profiles along the first- and second-nearest neighbor axes are shown in Figure~\ref{supp:fig:density_hex} in the SI, the corresponding free energy profiles are given in Figure~\ref{fig:free_energy_hex}.

The free energy barriers for diffusion of lithium ions on the surface of the negative electrode amount to approximately $1.8$~$k_B T$ for transitions between next nearest neighbor sites and $2.0$~$k_B T$ for transitions between second next nearest neighbor sites. This is about $3.5$ times smaller than the barrier for transitions from the second to the first \Lip{}-layer ($6.4$~$k_BT$) and about $4.5$ times smaller than the barrier for transitions from the first to the second \Lip{}-layer ($8.4$~$k_BT$). Hence, the lithium-ion transport on the electrode surface must be substantially faster than the lithium-ion transport towards or away from the electrode.

The free energy barrier for transitions between next nearest neighbors is slightly smaller than for transitions between second next nearest neighbors, because the lithium ions only have to cross a C--C bond instead of two carbon nuclei. The crossing of two carbon nuclei is also the reason for the double hump structure of the free energy profile along the second nearest neighbor axes. Given the higher energy barrier for transitions between second next nearest neighbors, the fact that the pathway between second nearest neighbors is longer than between nearest neighbors and the finding that the lithium-ion transport on the electrode surface must be faster than the lithium-ion transport toward or away from the electrode, one expects that the lithium ions travel on the surface primarily via hopping between nearest neighbor sites (see also below).

\subsubsection*{Lithium-Ion Diffusion}

The emergence of free energy barriers leads to a reduction of the lithium-ion diffusion near the electrodes. We measured the lithium-ion diffusion by means of the variance $\text{Var}[\Delta z]$ of the displacement vectors $\Delta z$ of the lithium ions as function of their initial $z$-position. The displacement variance corresponds to the usually calculated mean-square displacement (MSD) $\langle \Delta z^2 \rangle$ in bulk simulations. We did not simply calculate the MSD here, because our simulation boxes are limited in $z$-direction by the electrodes. Thus, particles initially located near an electrode are restricted in their $z$-movement and likely accumulate a net-drift towards the bulk region. This net-drift artificially increases their MSD when calculating the MSD as function of their initial $z$-position. To account for this net-drift, we subtracted the square of the mean displacement, $\langle \Delta z \rangle^2$, from the mean-square displacement, $\langle \Delta z^2 \rangle$. Mathematically, this results in the variance of the distribution of particle displacements:
\begin{equation}
  \text{Var}[\Delta z] = \langle \Delta z^2 \rangle - \langle \Delta z \rangle^2
  \label{eq:displvar}
\end{equation}
The brackets $\langle ... \rangle$ denote averaging over all particles of the same type with the same initial $z$-position $z(t_0)$ and over multiple starting times $t_0$. The displacement $\Delta z$ after a lag time $\Delta t$ was calculated as
\begin{equation}
  \Delta z \equiv \Delta z(\Delta t) = z(t_0 + \Delta t) - z(t_0).
  \label{eq:displacement}
\end{equation}
For consistency reasons, we calculated the displacement variance instead of the MSD in $x$- and $y$-direction, too, although no significant net-drift in these two directions was observed (at least not at short diffusion times, see Figures~\ref{supp:fig:mean_displ_time} and \ref{supp:fig:mean_displ_cross_section_Li} in the SI).

\begin{figure*}[!ht]
  \centering
  \includegraphics[width=\textwidth]{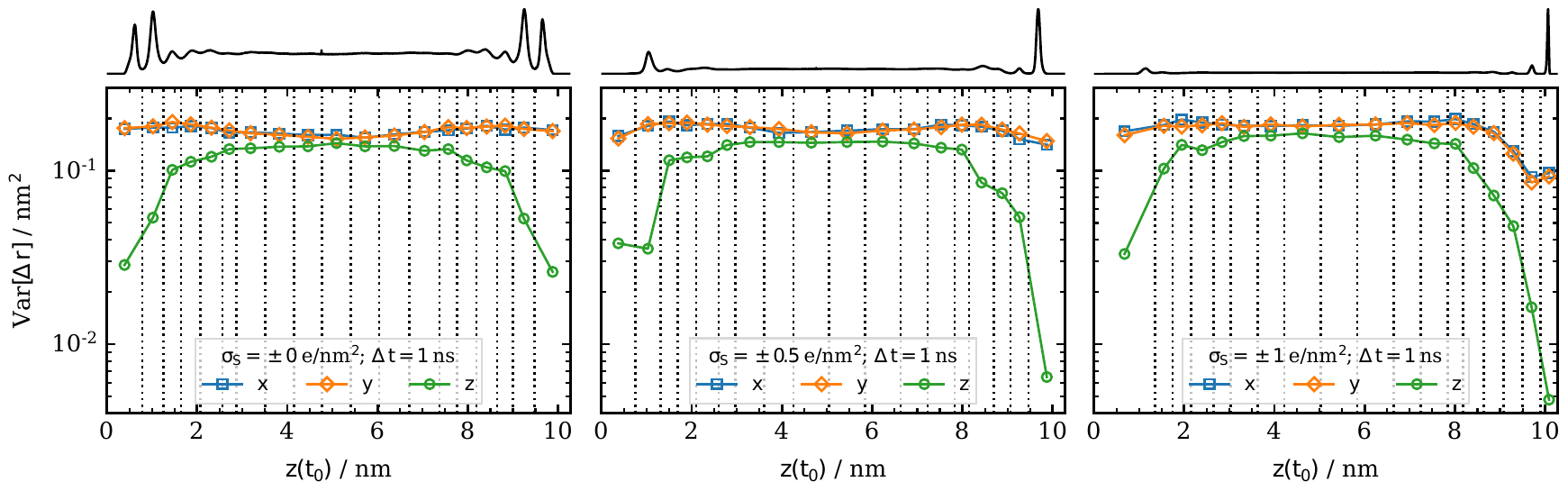}
  \caption{Displacement variance of lithium ions in $x$-, $y$- and $z$-direction as function of the initial $z$-position, evaluated at a lag time of \SI{1}{\nano\second} for the three investigated surface charges (left to right). The black curves on the top show the \Lip{}-density profiles. The vertical dotted lines indicate the discretization of the simulation boxes into bins.}
  \label{fig:displvar_cross_section}
\end{figure*}
To calculate the displacement variances as function of the initial $z$-position, the simulation boxes were discretized into bins in $z$-direction. Thereby the layer structure was taken into account such that the bins near the electrodes correspond to the \Lip{}-layers. The bins in the bulk region were chosen equidistant. The discretization scheme is shown in Figure~\ref{supp:fig:z-bins} in the SI and is additionally indicated in Figure~\ref{fig:displvar_cross_section} by the vertical dotted lines. All particles whose initial $z$-position is in the same bin are said to have the same initial $z$-position.

Figure~\ref{fig:displvar_cross_section} shows the displacement variances at a fixed lag time of \SI{1}{\nano\second} as a function of the initial $z$-position for all three spatial directions and all three investigated surface charges. The full displacement variances as function of time are shown in Figure~\ref{supp:fig:displvar_time} in the SI.

\paragraph*{Perpendicular to the Electrodes.}

For all three investigated surface charges, the lithium-ion diffusion perpendicular to the electrodes (i.e.\ in $z$-direction) is significantly reduced in the regions where layering occurs, as can be seen by the strong decrease of the $z$-component of the displacement variance. At uncharged surfaces, the $z$-displacement variance decreases by a factor of roughly five compared to the bulk region. At positive surfaces, the $z$-displacement variance decreases by approximately the same factor. The similar decrease at uncharged and positive surfaces can be attributed to the similar free energy barriers near these surfaces, which lead to similar mean residence times of lithium ions in \Lip{}-layers near these surfaces (see Table~\ref{supp:tab:lifetime_layer} in the SI). At negative surfaces, the $z$-displacement variance decreases even more by factors of roughly $20$ in the case of $\sigma_S = -\SI{0.5}{\elementarycharge\per\square\nano\meter}$ and $25$ in the case of $\sigma_S = -\SI{1}{\elementarycharge\per\square\nano\meter}$. This is in qualitative agreement with the higher free energy barriers and mean residence times (see again Table~\ref{supp:tab:lifetime_layer} in the SI) near negative surfaces. The reduced $z$-displacement variances in the regions where layering occurs reveal that the layer structure impedes the perpendicular \Lip{}-transport and therefore gives rise to additional resistances.

The $z$-displacement variances additionally show that a true bulk regime is not present in neither of the three surface simulations. Although the layer structure is only visible up to \SIrange{3}{3.5}{\nano\meter} from the surface, there is still a slight increase of the $z$-displacement variances when approaching the center of the simulation boxes at $\SI{5.14}{\nano\meter}$. Additionally, the $z$-displacement variances are always less then the $x$- and $y$-displacement variances, although they should be the same in a true bulk regime due to the isotropic nature of bulk PEO-\LiTFSI{} mixtures. Thus, besides the layering effect also more subtle structural effects are present which extend beyond $\SI{3.5}{\nano\meter}$ from the surface and affect the lithium-ion dynamics. The term \enquote{bulk} is therefore not justified for this region, rather it should be called \enquote{pseudo-bulk}. However, for the sake of consistency, we will still call it \enquote{bulk} region in the remainder of this paper.

\paragraph*{Parallel to the Electrodes.}

The displacement variance parallel to the electrodes (i.e.\ in $x$- and $y$-direction) does not show such a strong dependence on the distance to the electrodes like the $z$-displacement variance. Surprisingly, when approaching the electrodes from the bulk region, the $x$- and $y$-displacement variances first increase slightly in the regions where layering occurs. This suggests that the lithium-ion diffusion is slightly faster inside the layers. In the first and second \Lip{}-layers at uncharged and positive surfaces, the parallel \Lip{}-displacement variance decreases again marginally. This decrease is likely caused by lithium ions that coordinate to anions or polymer segments that are immobilized on the surface. At the highly negative surface with $\sigma_S = -\SI{1}{\elementarycharge\per\square\nano\meter}$, where the lithium ions get immobilized themselves by coordinating directly to the surface, the parallel displacement variance decreases by a factor of approximately two. This immobilization can be attributed to the free energy barriers for diffusion of lithium ions on the electrode surface.

Interestingly, the parallel displacement variance in the second \Lip{}-layer at the electrode with $\sigma_S = -\SI{1}{\elementarycharge\per\square\nano\meter}$ is almost the same as in the first layer, although no internal structure indicating free energy barriers within the second layer was found. However, the immobilized lithium ions in the first layer slow down the segmental dynamics of the polymer chains that coordinate to these lithium ions significantly, as indicated by the decreased displacement variance of the ether oxygens of PEO at the negative electrode in Figure~\ref{supp:fig:displvar_cross_section_OE}b in the SI. The reduced segmental dynamics in turn slows down the lithium ions in the following layers.

\paragraph*{On the Electrode Surface.}

\begin{figure}[!ht]
  \centering
  \includegraphics[width=\columnwidth]{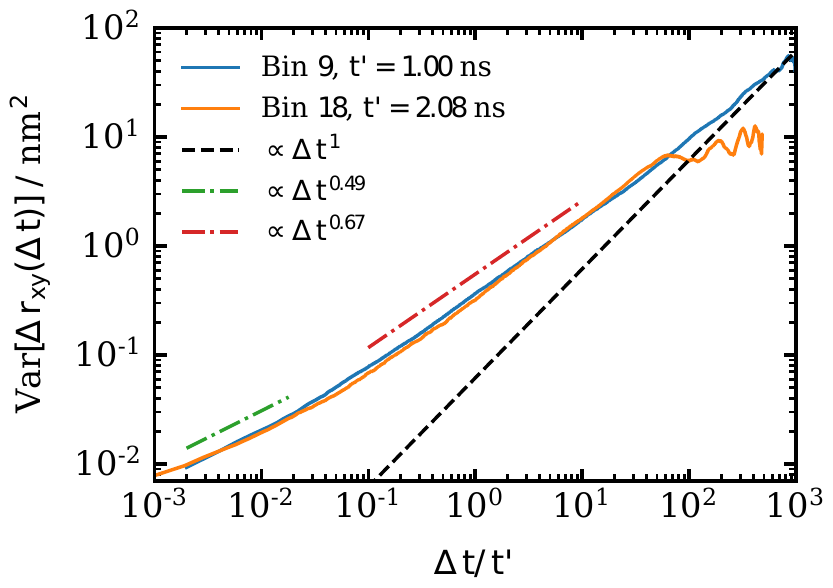}
  \caption{Time-scaled parallel \Lip{}-displacement variance (sum of $x$- and $y$-direction) for lithium ions initially located in the center of the bulk region (Bin~$9$) or in the first layer at the negative electrode with $\sigma_S = -\SI{1}{\elementarycharge\per\square\nano\meter}$ (Bin~$18$). The black dashed line indicates the diffusive regime. The dash-dotted green and red lines are obtained from fits of the parallel displacement variance in the center of the bulk region in the intervals \SIrange[range-phrase={ to }]{2e-3}{2e-2}{\nano\second} and \SIrange[range-phrase={ to }]{1e-1}{1e+1}{\nano\second}, respectively, with the fit function $v(\Delta t) = a \Delta t^\alpha$.}
  \label{fig:displvar_time_tscaled}
\end{figure}
The lithium-ion diffusion on the electrode surface is governed by the same transport mechanisms as in the bulk region. This can be inferred from the same slope of the parallel \Lip{}-displacement variance plotted as function of the diffusion time for lithium ions initially located in the first \Lip{}-layer at the negative electrode with $\sigma_S = -\SI{1}{\elementarycharge\per\square\nano\meter}$ and for lithium ions initially located in the center of the bulk region. In fact, the displacement variances can be brought into agreement by simply scaling the diffusion time, as shown in Figure~\ref{fig:displvar_time_tscaled}. Because the average continuous residence time of lithium ions in the first layer is \SI{13}{\nano\second} (see Table~\ref{supp:tab:lifetime_layer} in the SI), this agreement is not caused by an exchange of lithium ions between the first layer and the bulk, at least not at short to moderate diffusion times. Instead, the displacement variances of lithium ions in the two different regions obey the same power law $v(\Delta t) \propto \Delta t^\alpha$, just with different proportionality constants. The proportionality constant characterizes the particle mobility, whereas the exponent $\alpha$ characterizes the mechanism of the particle movement \cite{Metzler2004}. For $\alpha = 1$ we obtain diffusive motion, i.e.\ a random walk of the particles. Values of $\alpha$ below unity indicate subdiffusive motion. Because the parallel displacement variance of lithium ions at the electrode surface follows the same power law as in the bulk region, we conclude that the same transport mechanisms apply in both regions. The subdiffusive motion indicates that the lithium ions undergo a significant number of back jumps instead of performing a memoryless random walk between the distinct lithium sites on the electrode surface. This view is supported by the huge difference in the average continuous and discontinuous residence times a lithium ion stays at a given site on the electrode surface ($\tau_h^{con} = \SI{0.01}{\nano\second}$ and $\tau_h^{dis} = \SI{0.56}{\nano\second}$, see the section \enquote{Residence times} in the SI). Probably, the lithium ions are frequently pulled back by the polymer chains when they change their lattice site.

\subsubsection*{Role of Interchain Transfers for Layer Crossing}

Lithium-ion transport in polymer electrolytes occurs basically via three mechanisms: i) diffusion along the polymer chain, ii) movement together with polymer segments and iii) transfer between different polymer chains \cite{Borodin2006, Maitra2007}. Interchain transfers are of key relevance for long-range \Lip{}-transport, because they open up new pathways for the lithium ions. Without interchain transfers, a given lithium ion could only explore the finite space determined by the polymer chain it coordinates to \cite{Borodin2006, Maitra2007}. Although the crossing of a free energy barrier between two \Lip{}-layers is a short-range transport, it is of particular interest to see whether interchain transfers are required for these crossing events, i.e. whether they are of great importance for the perpendicular lithium-ion transport near interfaces, too.

\begin{figure}[!ht]
  \centering
  \includegraphics[width=\columnwidth]{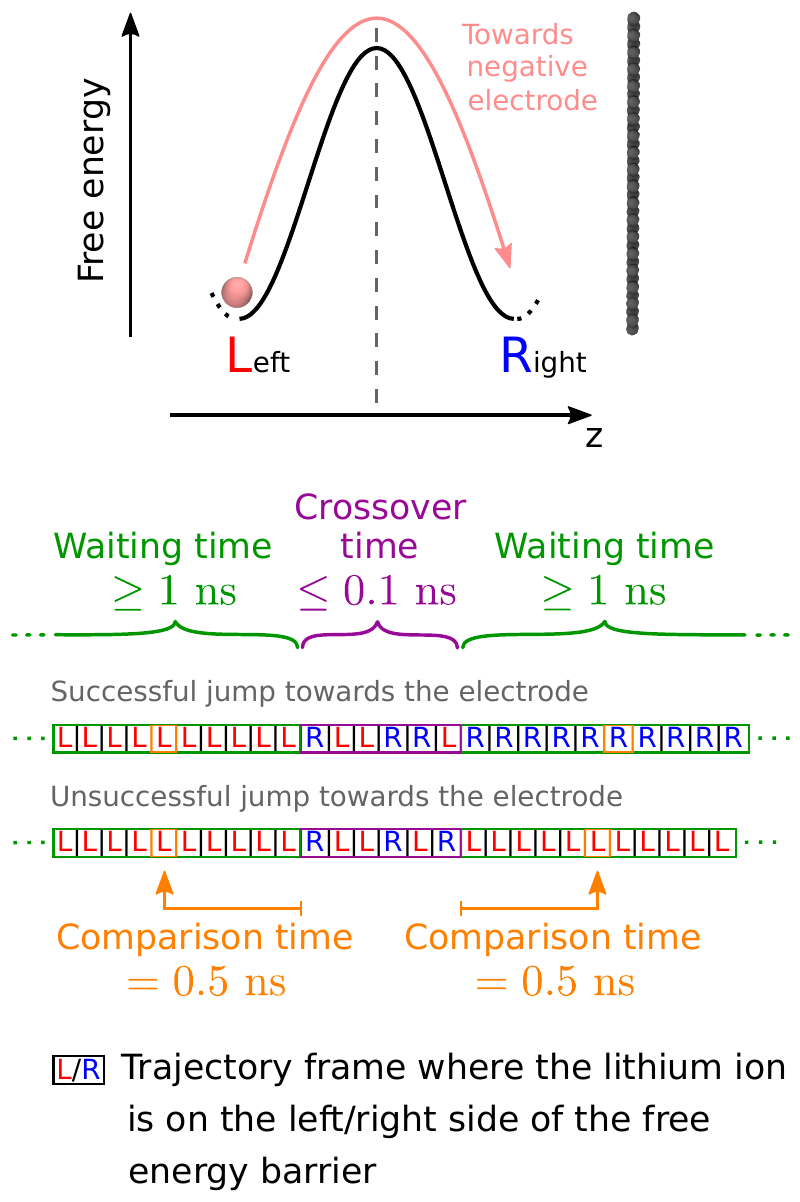}
  \caption{Setup to compare the coordination environment of a lithium ion before and after it has crossed an energy barrier. Only crossing events that took not more than $\SI{0.1}{\nano\second}$ and where the lithium ion stayed continuously at least $\SI{1}{\nano\second}$ on one side of the barrier before and after the jump were considered. The coordination environment of the lithium ion is compared $\SI{0.5}{\nano\second}$ before and after the crossing event.}
  \label{fig:lig_change_at_pos_change_definitions}
\end{figure}
\begin{figure*}[!ht]
  \centering
  \includegraphics[width=\textwidth]{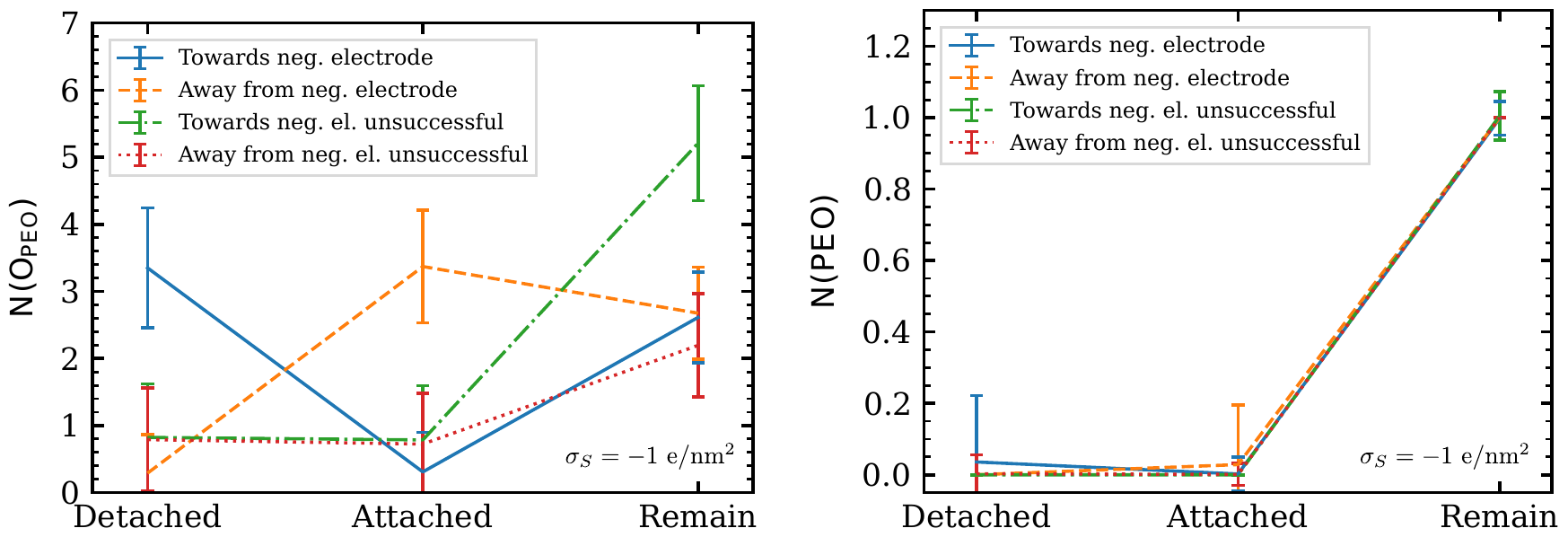}
  \caption{Exchange of ether oxygens from PEO (left) and of entire PEO chains (right) that coordinate to lithium ions that cross the free energy barrier between the first and second \Lip{}-layer ($z = \SI{9.9}{\nano\meter}$) at the electrode with $\sigma_S = -\SI{1}{\elementarycharge\per\square\nano\meter}$. The error bars indicate the standard deviation of the coordination numbers. The corresponding plots for $\sigma_S = 0$ and $-\SI{0.5}{\elementarycharge\per\square\nano\meter}$ are shown in Figure~\ref{supp:fig:lig_change_at_pos_change_near_negative_electrode} in the SI.}
  \label{fig:lig_change_at_pos_change}
\end{figure*}
To answer this question, we compared the coordination environment of lithium ions before and after they have crossed an energy barrier between two \Lip{}-layers, as sketched in Figure~\ref{fig:lig_change_at_pos_change_definitions}. Like above, an oxygen atom is said to be coordinated to a lithium ion, when it is within a \SI{0.3}{\nano\meter} cutoff radius of the ion. To dismiss immediate backjumps, we only considered lithium ions that stayed continuously for at least $\SI{1}{\nano\second}$ on one side of the energy barrier and after a crossover time stayed again continuously for at least $\SI{1}{\nano\second}$ on the other or the same side of the barrier. This $\SI{1}{\nano\second}$ waiting time corresponds to the average continuous residence times of lithium ions in the first layer at uncharged and positive electrodes (see Table~\ref{supp:tab:lifetime_layer} in the SI). The crossover time was allowed to be no longer than $\SI{0.1}{\nano\second}$. The actually counted layer crossings usually took \SIrange{0.01}{0.06}{\nano\second} (see Table~\ref{supp:tab:lig_change_at_pos_change} in the SI). To give the lithium ions enough time to change their polymer chain, we compared the coordination environment of the lithium ions \SI{0.5}{\nano\second} before the layer crossing with the coordination environment \SI{0.5}{\nano\second} afterwards. The crossing events can be classified into four categories: i) jumps towards the negative (or positive) electrode, ii) jumps away from the negative (or positive) electrode, iii) unsuccessful jumps towards the electrode and iv) unsuccessful jumps away from the electrode. Here, unsuccessful means that after being at least $\SI{1}{\nano\second}$ at one side of the energy barrier, the lithium ion jumps to the other side but jumps back to its original side within the $\SI{0.1}{\nano\second}$ crossover time and stays there for at least another nanosecond.

Figure~\ref{fig:lig_change_at_pos_change} shows how the coordination environment of lithium ions changes that cross the energy barrier at $z = \SI{9.9}{\nano\meter}$ in the system with $\sigma_S = \pm \SI{1}{\elementarycharge\per\square\nano\meter}$, i.e.\ the energy barrier between the first and second layer at the negative electrode. Similar plots for the systems with $\sigma_S = 0$ and $\pm \SI{0.5}{\elementarycharge\per\square\nano\meter}$ and for the first energy barrier at positive electrodes are shown in Figures~\ref{supp:fig:lig_change_at_pos_change_near_negative_electrode} and \ref{supp:fig:lig_change_at_pos_change_near_positive_electrode} in the SI.

Lithium ions that jump towards the negative electrode detach on average $3.4$ of their coordinating ether oxygens but attach almost no new ether oxygens in turn. This is due to the partial desolvation of lithium ions that come into direct contact with the negative electrode. Lithium ions that move away from the electrode surface attach $3.4$ ether oxygens. However, in either case always $2.6$ of the original ether oxygens remain coordinated to the lithium ions, indicating that a lithium ion never leaves its polymer chain. The right panel of Figure~\ref{fig:lig_change_at_pos_change} confirms that indeed the same polymer chain remains coordinated to the lithium ion during the layer crossing. Equivalent results were obtained for the systems with other surface charges and for layer crossings near positive electrodes (Figures~\ref{supp:fig:lig_change_at_pos_change_near_negative_electrode} and \ref{supp:fig:lig_change_at_pos_change_near_positive_electrode} in the SI). This clearly shows that interchain transfers are not necessary for the lithium ions to jump between different \Lip{}-layers and in fact only play a minor role in layer crossing processes.

The left panel of Figure~\ref{fig:lig_change_at_pos_change} shows that in the case of unsuccessful layer crossing events on average only $0.8$ ether oxygens are exchanged, but $5.2$ and $2.2$ of the original ether oxygens remain coordinated in the case of unsuccessful jumps towards and away from the electrode, respectively. Although the lithium ions do not change their polymer chain to cross the free energy barrier between the first and second layer, they have to get either partially desolvated or solvated (depending on the jump direction) in order to successfully cross this barrier. At other free energy barriers, lithium ions must partially exchange their coordination spheres to cross the barrier (see Figures~\ref{supp:fig:lig_change_at_pos_change_near_negative_electrode} and \ref{supp:fig:lig_change_at_pos_change_near_positive_electrode} in the SI). However, at these barriers the exchange of the coordination sphere is not noticeably different between successful and unsuccessful jumps. It seems like lithium ions simply move along a polymer chain or follow its segmental motion and by chance thereby transition between different \Lip{}-layers.

At almost all inspected free energy barriers, the number of unsuccessful crossing events is at least an order of magnitude larger than the number of successful events (see Table~\ref{supp:tab:lig_change_at_pos_change} in the SI). This indicates that the lithium ions undergo a significant number of backjumps to their original layer before finally leaving it. Note that immediate backjumps are even not incorporated in these numbers due to the $\SI{1}{\nano\second}$ waiting time criterion. We have seen a similar behavior for the movement of lithium ions between the different hexagonal sites on the negative electrode with $\sigma_S = -\SI{1}{\elementarycharge\per\square\nano\meter}$ (see above). Again, the backjumps are also reflected in the continuous and discontinuous probabilities that a lithium ion stays in a given \Lip{}-layer (Figure~\ref{supp:fig:lifetime_layer} in the SI). As outlined in the SI, the discontinuous probabilities decay at least an order of magnitude later than the continuous probabilities, indicating that lithium ions frequently jump back into their original layer. A likely explanation for the frequent backjumps is that the lithium ions are pulled back by the polymer chains. Only jumps from the second to the first layer at the negative electrode with $\sigma_S = -\SI{1}{\elementarycharge\per\square\nano\meter}$ are more often successful than unsuccessful ($446$ successful versus $215$ unsuccessful jumps). Here, the strong electrostatic attraction that the lithium ions experience once they are directly attached to the negative surface counteracts the reset force of the polymer chains.

\section*{Conclusions}

We have investigated the atomistic structure and lithium-ion dynamics of a PEO-\LiTFSI{} polymer electrolyte in the vicinity of uncharged and charged model electrodes with molecular dynamics simulations. The presence of surfaces induces a layering of the electrolyte components. The shape and density of the layers heavily depend on the surface charge. Only at high surface charges of $\sigma_S = -\SI{1}{\elementarycharge\per\square\nano\meter}$, lithium ions are released from their tight PEO solvation cage and come into direct contact with the electrode. Lithium ions that are directly attached to the negative electrode favor to reside at the center of the graphene hexagons. Common for all surface charges is the observation that \Lip{}-layers are almost always surrounded by two \OPEO{}-layers, emphasizing the strong Li-PEO interaction. Only near the positive electrode, lithium ions are also coordinated by \TFSI{} anions.

The structural effects close to interfaces have direct implications for the dynamics. For the lithium-ion dynamics parallel to the surfaces, these implications are small. Only when the lithium ions are directly attached to the electrode, the parallel lithium-ion diffusion is reduced by a factor of two. But despite the emergence of distinct lithium sites at the surface, the lithium-ion dynamics on the surface is still governed by polymer effects like in the bulk. Specifically, the lithium ions frequently jump back to their original hexagon sites, probably because they are pulled back by the polymer chains. The surface itself only leads to a general decrease of the parallel lithium-ion dynamics, but does not affect the transport mechanism.

Perpendicular to the surfaces, the lithium-ion dynamics is changed more drastically. The layering effect, going along with the emergence of free energy barriers between the layers, substantially reduces the lithium-ion diffusion perpendicular to the surfaces. This effect is strongest close to negatively charged surfaces. At positively charged surfaces, the reduction of the lithium-ion diffusion is almost the same as at uncharged surfaces. In terms of battery applications, the layering gives rise to interface resistances at the electrodes that must be overcome by the lithium ions. Similar to the movement on the electrode surface, the lithium ions often jump back to their initial layers, which is probably because the polymer chains apply a reset force on the ions. However, a comparison of the coordination environment of lithium ions before and after layer crossing revealed that interchain transfers play no significant role for transitions between different \Lip{}-layers.

\begin{suppinfo}
  Supporting Information: Density profiles of all electrolyte components, $z$-dependent lithium-ion coordination numbers for the systems with $\sigma_S = 0$ and $\pm\SI{0.5}{\elementarycharge\per\square\nano\meter}$, $xy$-density maps of \OBT{} and \OPEO{} at the positive electrode with $\sigma_S = +\SI{1}{\elementarycharge\per\square\nano\meter}$, \Lip{}-density profiles along the first- and second-nearest neighbor axes as indicated in Figure~\ref{fig:densmap-xy_negative_electrode}a, a plot illustrating the partition of the simulation boxes into $z$-bins, mean displacements and displacement variances of \Lip{} as function of diffusion time, mean displacement of \Lip{} as function of initial $z$-position, mean displacement and displacement variance of \OPEO{} as function of initial $z$-position, mean residence times of lithium ions in the different \Lip{}-layers and on the distinct hexagonal sites as indicated in Figure~\ref{fig:densmap-xy_negative_electrode}a, plots illustrating the changes in the Li-polymer coordination when crossing different energy barriers, a table containing the number and crossover times of counted crossing events, force field validation, including densities, radial distribution functions, lithium-ion coordination numbers, mean-square displacements and diffusion coefficients.
\end{suppinfo}

\begin{acknowledgement}
  The authors thank the German Federal Ministry of Education and Research for financial support through the FestBatt project (grant number 03XP0174B). Parts of the analysis computations were performed on the PALMA-II HPC cluster of the University of Münster.
\end{acknowledgement}

\putbib[literature_manuscript_bibtex]
\end{bibunit}

\clearpage
\newgeometry{margin=2.54cm}  % ACS draft layout
\onecolumn
\doublespacing

\section*{Graphical TOC Entry}
\begin{figure}[H]
  \centering
  \includegraphics{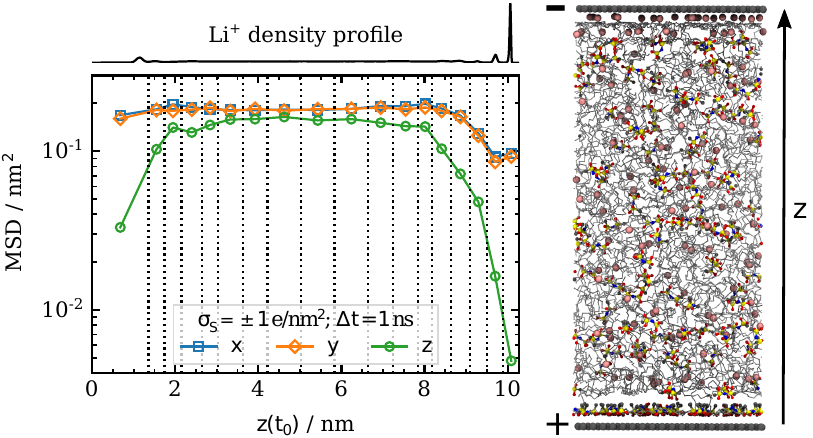}
  \label{toc_graphic}
\end{figure}

\cleardoublepage
\pagestyle{plain}
\setcounter{page}{1}
\setcounter{figure}{0}
\setcounter{table}{0}
\setcounter{equation}{0}
\renewcommand{\thepage}{S\arabic{page}}
\renewcommand{\thefigure}{S\arabic{figure}}
\renewcommand{\thetable}{S\arabic{table}}
\renewcommand{\theequation}{S\arabic{equation}}

\begin{bibunit}
\begin{center}
  \begin{LARGE}
    \bfseries
    \sffamily
    Supporting Information for\\
    \vspace{\baselineskip}
    Impact of Charged Surfaces on the Structure and Dynamics of Polymer Electrolytes: Insights from Atomistic Simulations\\
    \vspace{0.4\baselineskip}
  \end{LARGE}
  \begin{large}
    \sffamily
    Andreas Thum,$^{*,\dagger}$ Diddo Diddens,$^{*,\ddagger}$ and Andreas Heuer$^{*,\dagger,\ddagger}$\\
    \vspace{0.3\baselineskip}
  \end{large}
  {
    \itshape
    $\dagger$Universität Münster, Institut für Physikalische Chemie, Corrensstraße 28/30, 48149 Münster, Germany\\
    $\ddagger$Helmholtz-Institut Münster: Ionenleitung in der Energiespeicherung (IEK-12), Forschungszentrum Jülich GmbH, Corrensstraße 46, 48149 Münster, Germany\\
    \vspace{0.3\baselineskip}
  }
  {
    \sffamily
    \urlstyle{sf}
    E-Mail: \url{a.thum@uni-muenster.de}; \url{d.diddens@fz-juelich.de}; \url{andheuer@uni-muenster.de}
  }
\end{center}

\section*{Electrolyte Structure}
\subsection*{Layering}

\begin{figure}[H]
  \centering
  \includegraphics[width=\textwidth]{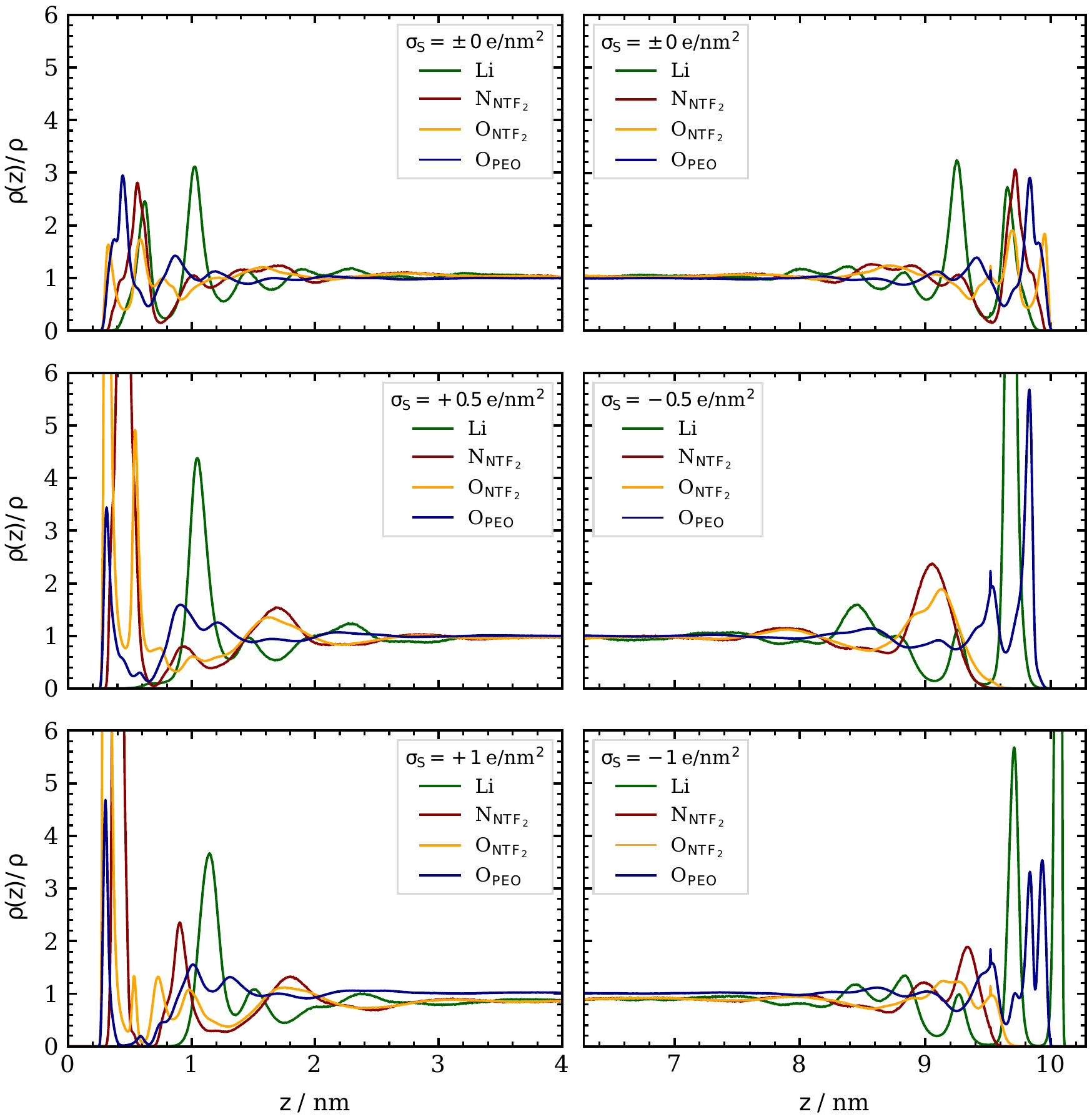}
  \caption{Normalized density profiles of \Lip{}, the nitrogen and oxygen atoms of \TFSI{} and the oxygen atoms of PEO perpendicular to the electrodes (i.e.\ in $z$-direction) for the three investigated surface charges $\sigma_S = 0$, $\pm 0.5$ and $\pm \SI{1}{\elementarycharge\per\square\nano\meter}$ (top to bottom). The positive electrode resides at $z = \SI{0}{\nano\meter}$ and the negative one at $z = \SI{10.28}{\nano\meter}$, which is the right boundary of the right plots. The density profiles are normalized by the bulk density such that a value of $1$ indicates the same density as in the bulk.}
  \label{supp:fig:dens-z}
\end{figure}

\subsection*{Lithium-Ion Coordination}

\begin{figure}[H]
  \centering
  \includegraphics[width=\textwidth]{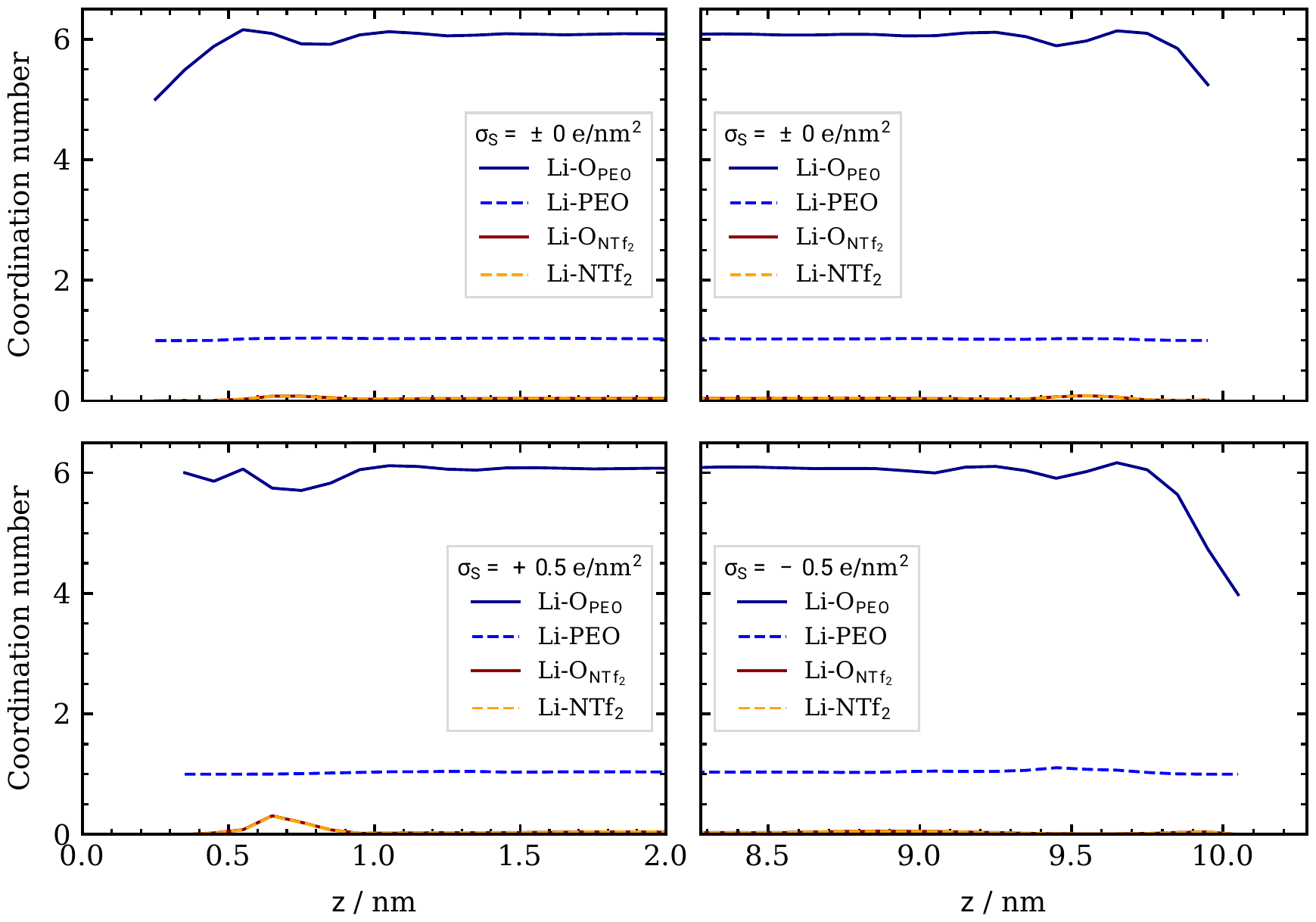}
  \caption{Average lithium-ion coordination numbers as function of $z$ for $\sigma_S = 0$ (top) and $\sigma_S = \pm \SI{0.5}{\elementarycharge\per\square\nano\meter}$ (bottom). Li-\OBT{} denotes coordinations of \Lip{} to single oxygen atoms from \TFSI{}. Li-\TFSI{} denotes coordinations of \Lip{} to entire \TFSI{} anions. The definitions of Li-\OPEO{} and Li-PEO are analogous. The corresponding plots for $\sigma_S = \pm \SI{1}{\elementarycharge\per\square\nano\meter}$ are shown in Figure~\ref{fig:coordination_number_z} in the main paper.}
  \label{supp:fig:coordination_number_z}
\end{figure}

\subsection*{Internal Structure of the First Layer ($\sigma_S = +\SI{1}{\elementarycharge\per\square\nano\meter}$)}

\begin{figure}[H]
  \centering
  \includegraphics[width=\textwidth]{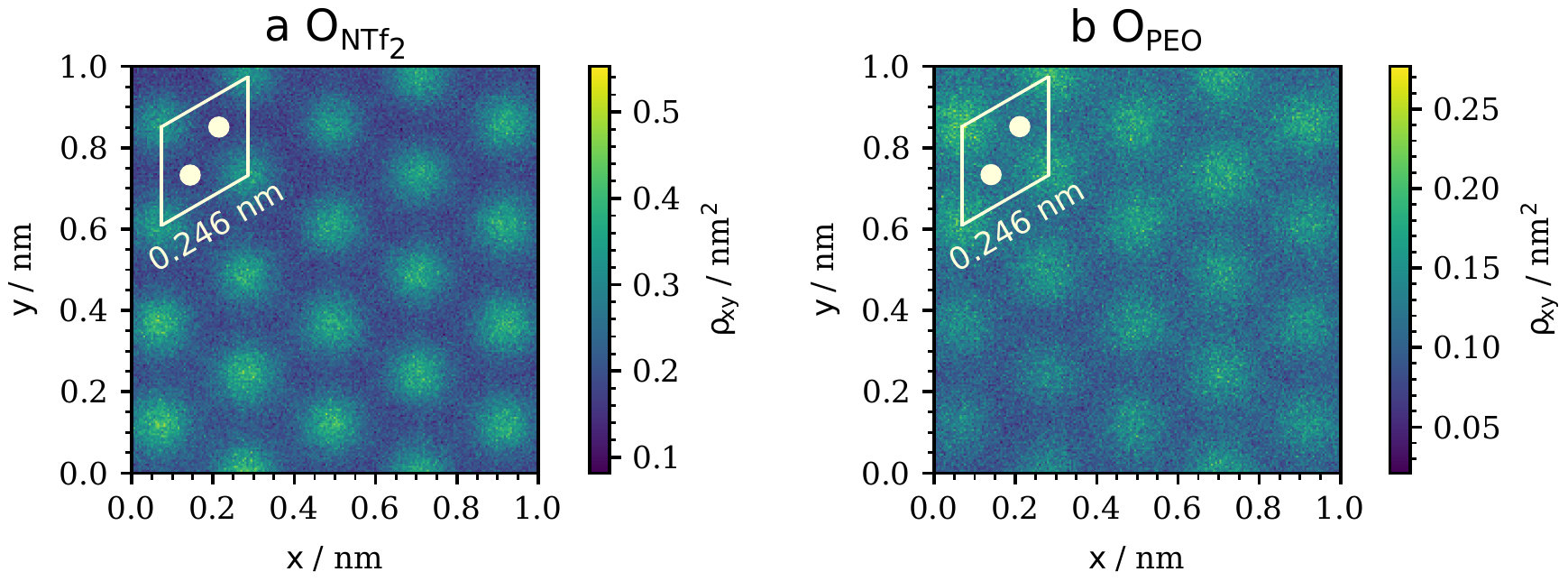}
  \caption{Density maps in $xy$-plane for (a) \OBT{} and (b) \OPEO{} in the first \OBT{}-layer ($z < \SI{0.50}{\nano\meter}$) and the first \OPEO{}-layer ($z < \SI{0.45}{\nano\meter}$) at the positive electrode with $\sigma_S = +\SI{1}{\elementarycharge\per\square\nano\meter}$. Because the patterns continue periodically, only the first square nanometer of the density maps is shown. In the upper left of both density maps, the unit cell of graphene is sketched in white.}
  \label{supp:fig:densmap-xy_positive_electrode}
\end{figure}

\section*{Charge Transport}
\subsection*{Free Energy Profiles}
\subsubsection*{On the Electrode Surface}

\begin{figure}[H]
  \centering
  \includegraphics{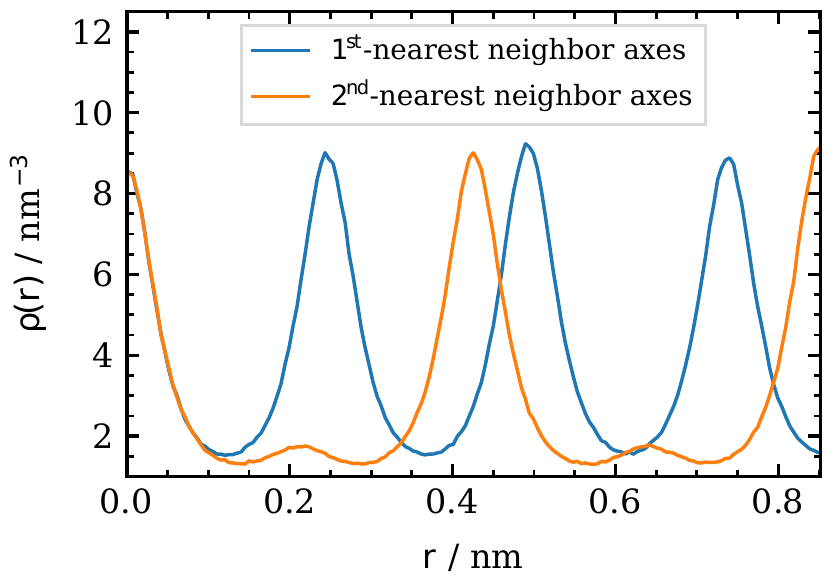}
  \caption{\Lip{}-density profiles along the first- and second-nearest neighbor axes of the hexagonal lithium sites in the first \Lip{}-layer ($z > \SI{9.9}{\nano\meter}$) at the negative electrode with $\sigma_S = -\SI{1}{\elementarycharge\per\square\nano\meter}$, as indicated in Figure~\ref{fig:densmap-xy_negative_electrode}a in the main paper. The shown profiles are the average over all first and second-nearest neighbor axes, respectively, and over all trajectory frames. Because the profiles continue periodically, only the first $\SI{0.852}{\nano\meter}$ are shown, which is six times the C-C bond length in graphene ($\SI{0.142}{\nano\meter}$). The origin $r = \SI{0}{\nano\meter}$ is set to the center of a graphene hexagon.}
  \label{supp:fig:density_hex}
\end{figure}

\clearpage
\subsection*{Lithium-Ion Diffusion}

\begin{figure}[H]
  \centering
  \includegraphics[width=\textwidth]{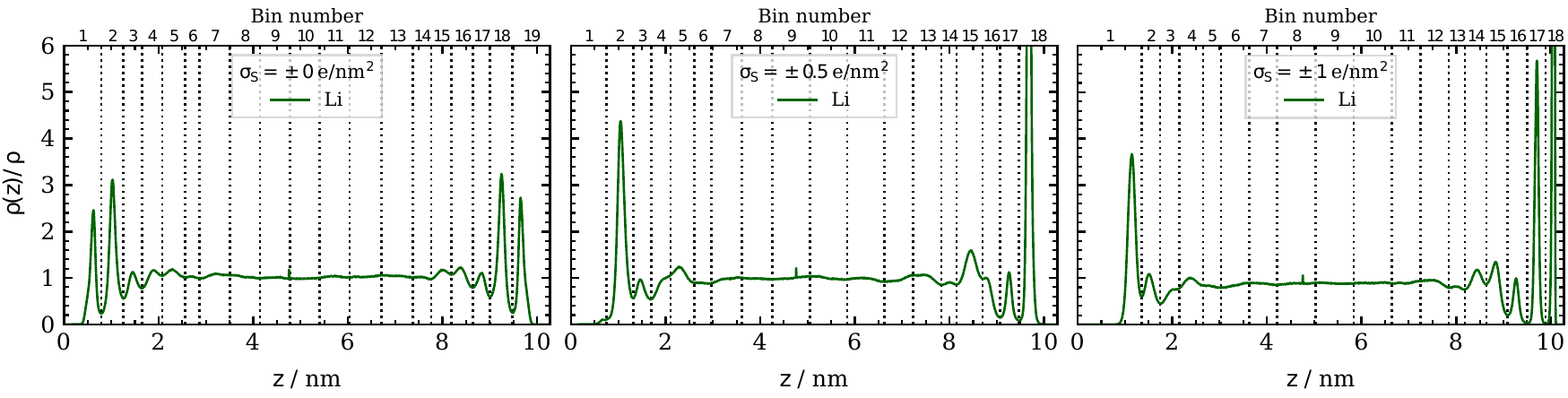}
  \caption{Normalized \Lip{}-density profiles showing the discretization of the simulation boxes into bins (vertical dotted lines) in $z$-direction for the three investigated surface charges (left to right). The bin numbers are shown on top.}
  \label{supp:fig:z-bins}
\end{figure}

\begin{figure}[H]
  \centering
  \includegraphics[width=\textwidth]{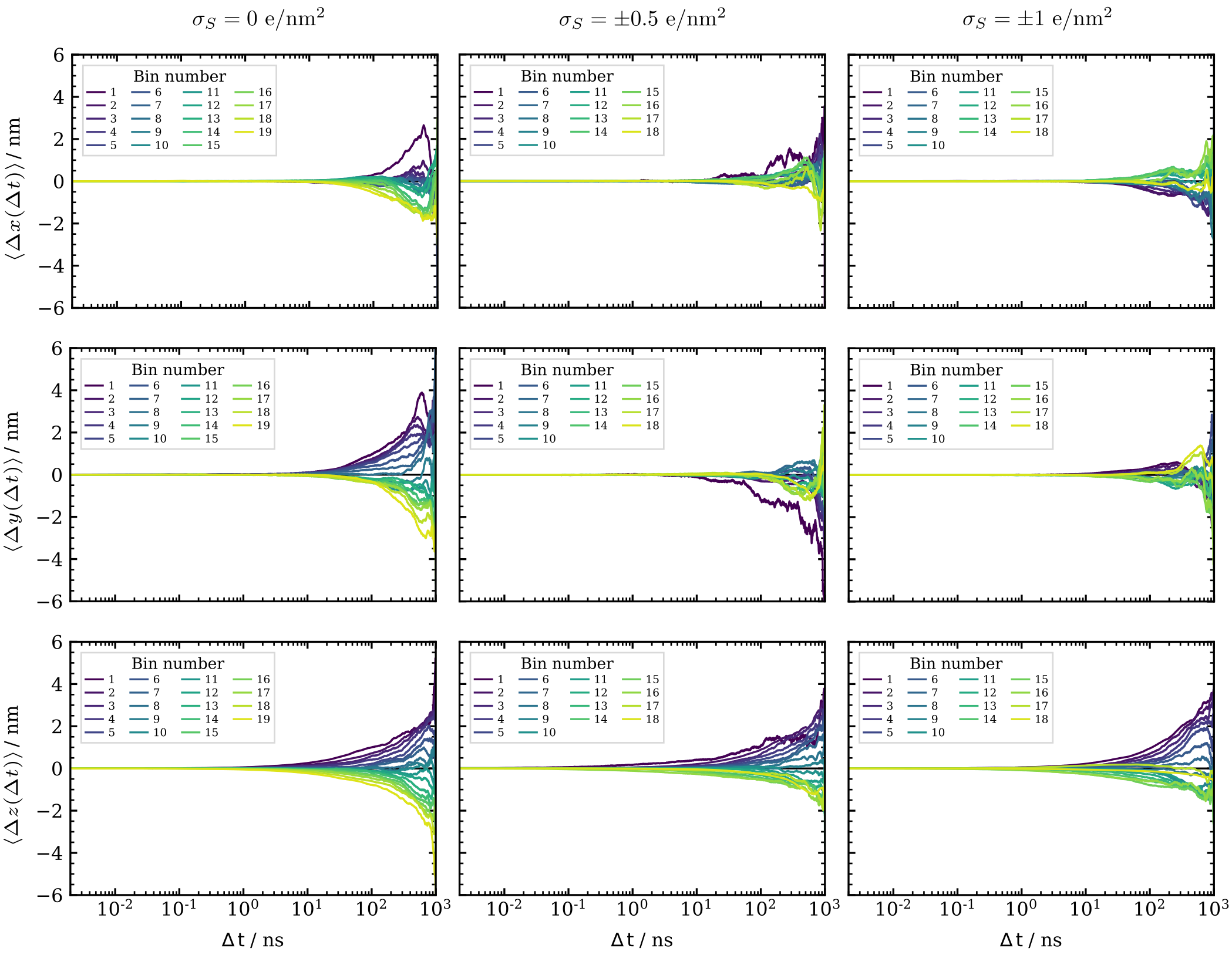}
  \caption{Mean displacements of lithium ions in $x$-, $y$- and $z$-direction (top to bottom) as function of diffusion time $\Delta t$, given that the lithium ions were initially located in a specific $z$-bin (see Figure~\ref{supp:fig:z-bins}) for the three investigated surface charges (left to right).}
  \label{supp:fig:mean_displ_time}
\end{figure}

\begin{figure}[H]
  \centering
  \includegraphics[width=\textwidth]{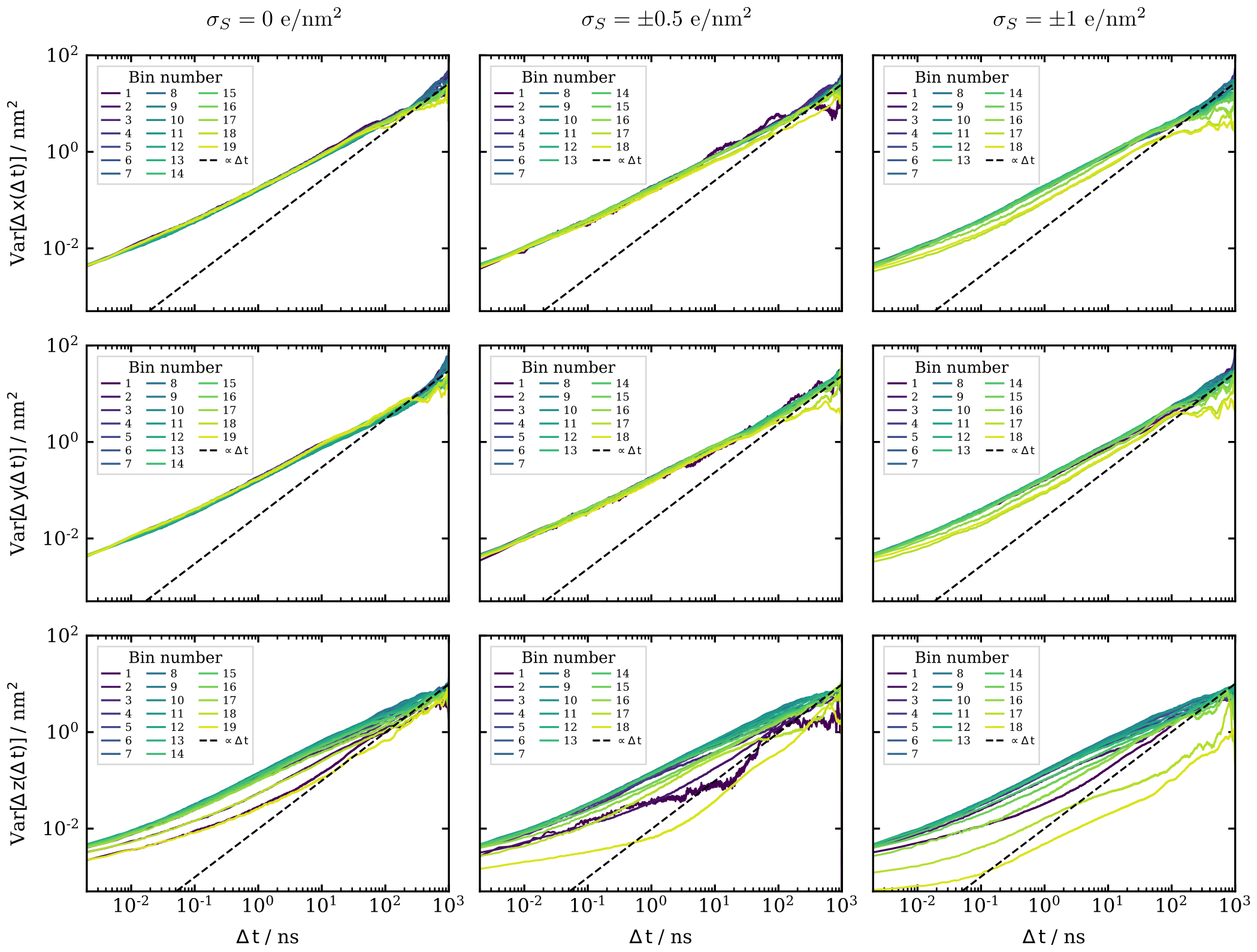}
  \caption{Displacement variances of lithium ions in $x$-, $y$- and $z$-direction (top to bottom) as function of diffusion time $\Delta t$, given that the lithium ions were initially located in a specific $z$-bin (see Figure~\ref{supp:fig:z-bins}) for the three investigated surface charges (left to right). The black dashed lines indicate the diffusive regime.}
  \label{supp:fig:displvar_time}
\end{figure}

\begin{figure}[H]
  \centering
  \includegraphics[width=\textwidth]{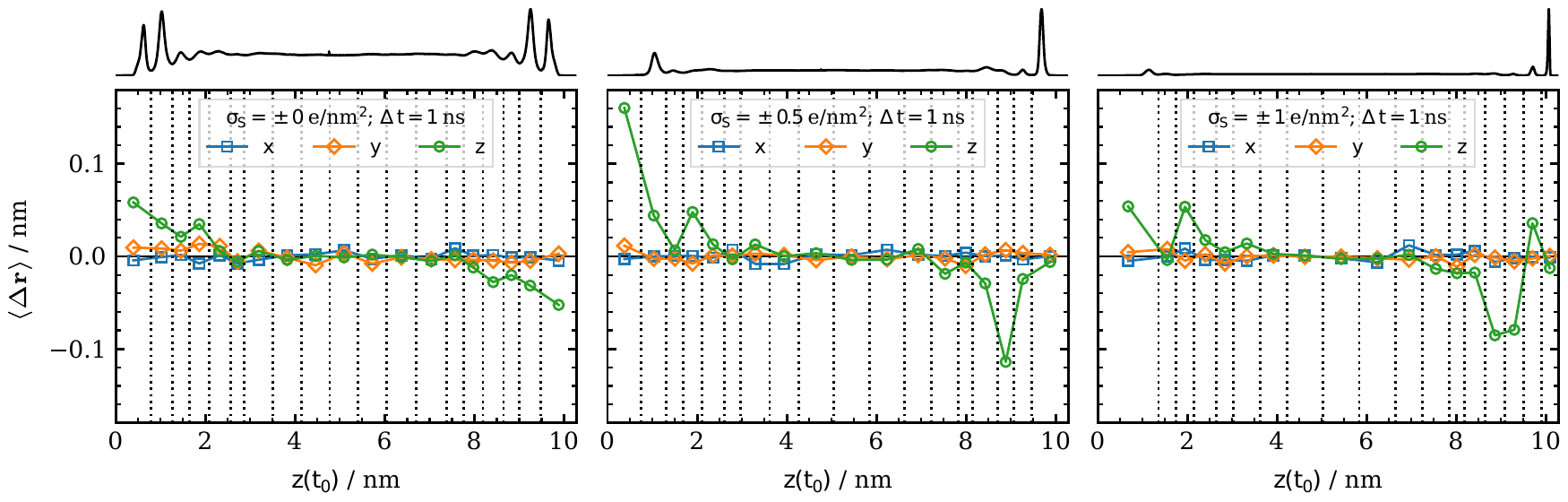}
  \caption{Mean displacement of lithium ions in $x$-, $y$- and $z$-direction as function of the initial $z$-position, evaluated at a lag time of \SI{1}{\nano\second} for the three investigated surface charges (left to right). The black curves on the top show the \Lip{}-density profiles. The vertical dotted lines indicate the discretization of the simulation boxes into bins.}
  \label{supp:fig:mean_displ_cross_section_Li}
\end{figure}

\begin{figure}[H]
  \centering
  \includegraphics[width=\textwidth]{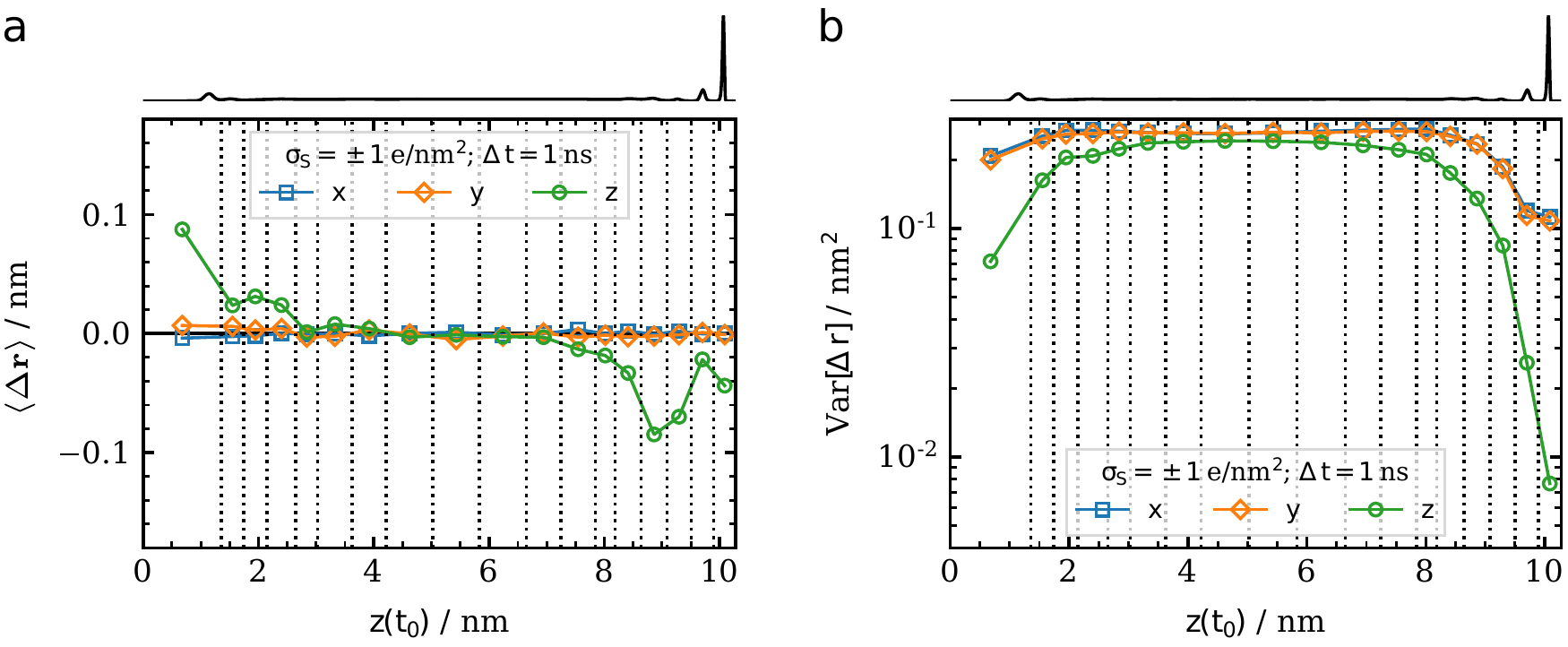}
  \caption{(a) Mean displacement and (b) displacement variance of PEO oxygens in $x$-, $y$- and $z$-direction as function of the initial $z$-position, evaluated at a lag time of \SI{1}{\nano\second} for the system with a surface charge of $\sigma_S = \pm \SI{1}{\elementarycharge\per\square\nano\meter}$. The black curves on the top show the \Lip{}-density profile. The vertical dashed lines indicate the discretization of the simulation box into bins.}
  \label{supp:fig:displvar_cross_section_OE}
\end{figure}

\clearpage
\subsection*{Residence Times}
\subsubsection*{Perpendicular to the Electrodes}

In the main paper we have shown that the lithium ions must overcome free energy barriers (Figure~\ref{fig:free_energy_z}) in order to transition between different \Lip{}-layers. Here, we estimate how long a lithium ion stays in a given layer before such a transition occurs.

The probability that a lithium ion is \textit{still} in the same \Lip{}-layer after a lag time $\Delta t$ is given by
\begin{equation}
p^{con}(\Delta t) \equiv p\left( z(t) \in L_i \ \forall \ t \in [t_0, \, t_0 + \Delta t] \right) = \left\langle \frac{ N_{L_i}\left( \left\{ t \in [t_0, \, t_0 + \Delta t] \right\} \right) }{N_{L_i}(t_0)} \right\rangle_t,
\label{supp:eq:lifetime_continuous}
\end{equation}
\makebox[\linewidth][s]{where $N_{L_i}(t_0)$ is the number of lithium ions that were in layer $L_i$ at time $t_0$ and}\\
$N_{L_i}\left( \left\{ t \in [t_0, \, t_0 + \Delta t] \right\} \right)$ is the number of lithium ions that have \textit{continuously} been in this layer \textit{at all times} from $t_0$ to $t_0 + \Delta t$. Therefore, we call this the continuous definition of the probability to stay in the same layer. The brackets $\langle ... \rangle_t$ denote the average over multiple starting times $t_0$ and $z(t)$ stands for the $z$-position of a given lithium ion at time $t$.

Besides the continuous definition of the probability, $p^{con}(\Delta t)$, also a discontinuous definition, $p^{dis}(\Delta t)$, can be made. The probability that a lithium ion is \textit{still or again} in the same \Lip{}-layer after a lag time $\Delta t$ is given by
\begin{equation}
p^{dis}(\Delta t) \equiv p\left( z(t_0 + \Delta t) \in L_i \mid z(t_0) \in L_i \right) = \left\langle \frac{N_{L_i}(t_0, \, t_0 + \Delta t)}{N_{L_i}(t_0)} \right\rangle_t.
\label{supp:eq:lifetime_discontinuous}
\end{equation}
Here, $N_{L_i}(t_0, \, t_0 + \Delta t)$ is the number of lithium ions that are \textit{still or again} in the same layer as they were at time $t_0$ after a lag time $\Delta t$. The discontinuous definition is sensitive to back jumps. If a lithium ion leaves the layer but comes back before $t_0 + \Delta t$, it is treated as it had never left the layer. Contrarily, the continuous definition treats it as ion that has left the layer.

\begin{figure}[!ht]
  \centering
  \includegraphics[width=\textwidth]{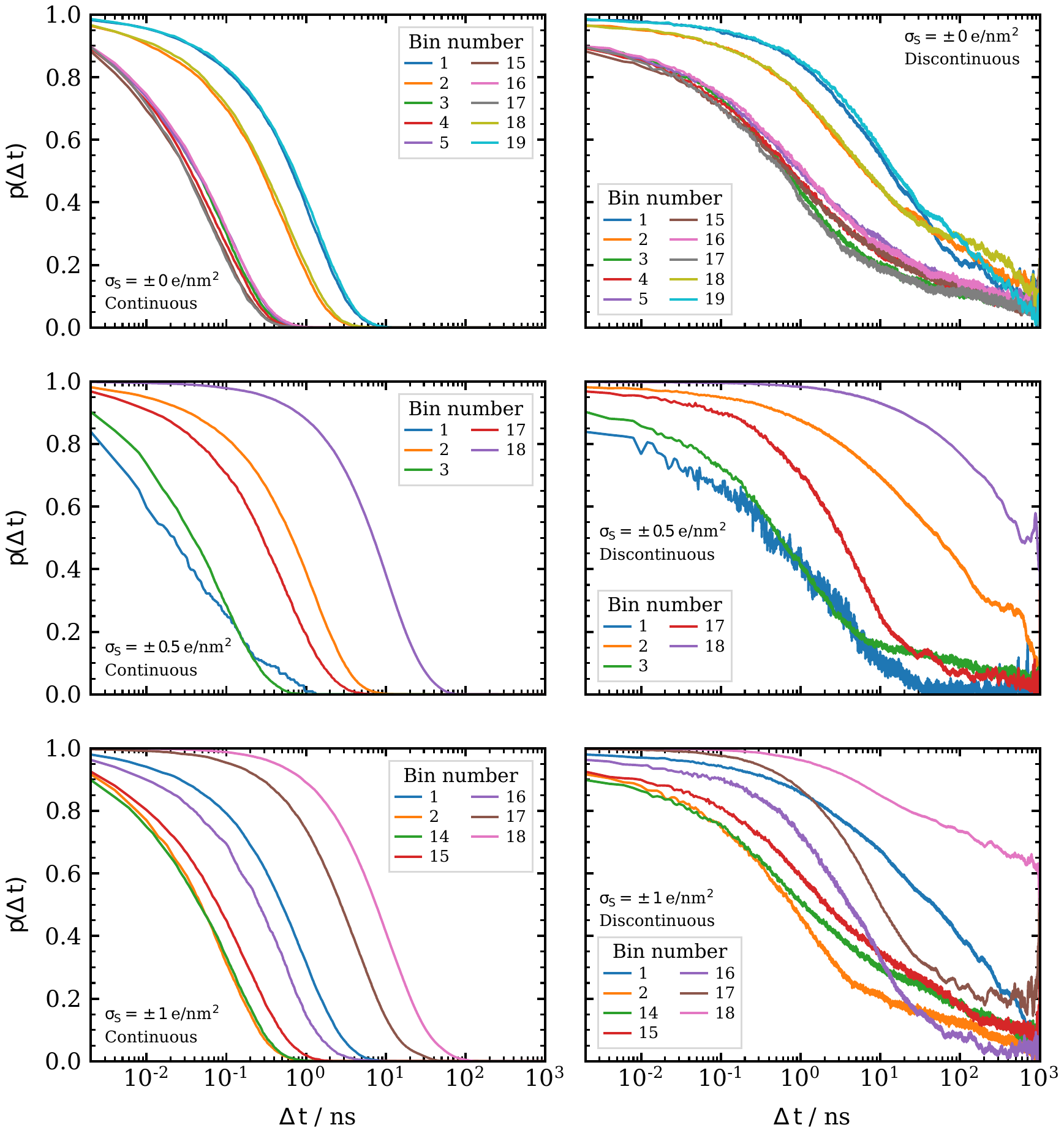}
  \caption{Probability that a lithium ion stays in a specific $z$-bin as function of lag time $\Delta t$ for bins that can be attributed to actual \Lip{}-layers (see Figure~\ref{supp:fig:z-bins}) for the three investigated surface charges (top to bottom). The left panel shows the results for the continuous definition of the probability, the right panel shows the results for the discontinuous definition (see text for more information).}
  \label{supp:fig:lifetime_layer}
\end{figure}

To calculate the probability for a lithium ion to be still (or again) in the same \Lip{}-layer, we used the same partition of the simulation box into bins as for calculating the $z$-dependent displacement variances (see Figure~\ref{supp:fig:z-bins}), but only focused on those bins that can be attributed to actual \Lip{}-layers. The progression of the continuous and discontinuous probability as function of lag time is shown in Figure~\ref{supp:fig:lifetime_layer} for the different layers of the different systems.

Relaxation processes in disordered systems with potentially multiple relaxation mechanisms are often described by a stretched exponential function, also known as Kohlrausch-Williams-Watts (KWW) function \cite{Kohlrausch1854, Williams1970, BerberanSantos2005, Johnston2006}:
\begin{equation}
f(t) = \exp{\left[-\left(\frac{t}{\tau_0}\right)^\beta\right]}
\label{supp:eq:kww}
\end{equation}
For physical systems, the stretching exponent $\beta$ is confined to $0 < \beta \leq 1$. The two parameters ($\beta$ and $\tau_0$) were determined for each \Lip{}-layer by fitting a stretched exponential to the probability curves until they decayed below $0.01$ or until $\SI{90}{\percent}$ of the maximum lag time of $\SI{1e3}{\nano\second}$ were reached (whatever happened earlier). The mean residence time $\tau_l$ of a lithium ion in a given layer can then be computed by integrating the stretched exponential from zero to infinity
\begin{equation}
\tau_l = \int_{0}^{\infty} f(t) \text{ d}t = \frac{t_0}{\beta} \ \Gamma\left(\frac{1}{\beta}\right),
\label{supp:eq:kww_integral}
\end{equation}
where $\Gamma(x)$ is the gamma function. For purely exponential decay ($\beta = 1$), $\tau_l = \tau_0$ applies.

From the discussion of $p^{con}(\Delta t)$ and $p^{dis}(\Delta t)$ above it follows that the mean residence time $\tau_l^{con}$ obtained from $p^{con}(\Delta t)$ is the average residence time a lithium ion continuously stays in a given \Lip{}-layer. After $\tau_l^{con}$ the lithium ion attempts to leave its layer. However, the layer transition might be unsuccessful in the sense that the lithium ion returns into its initial layer, for example because it is pulled back by the polymer. These back jumps are captured by $\tau_l^{dis}$ obtained from $p^{dis}(\Delta t)$. Therefore, $\tau_l^{dis}$ is usually greater than $\tau_l^{con}$. Only when no back jumps occur, $p^{dis}(\Delta t)$ and $p^{con}(\Delta t)$, and hence $\tau_l^{dis}$ and $\tau_l^{con}$, are equal.

Table~\ref{supp:tab:lifetime_layer} lists the average continuous residence times $\tau_l^{con}$ of lithium ions in $z$-bins that can be attributed to actual \Lip{}-layers. The average discontinuous residence times $\tau_l^{dis}$ are not given, because the discontinuous probabilities could not be fitted satisfactorily by stretched exponentials due to their slow decaying tails at long lag times. Nevertheless, the comparison of $p^{con}(\Delta t)$ and $p^{dis}(\Delta t)$ (see Figure~\ref{supp:fig:lifetime_layer}) reveals that in all cases $p^{dis}(\Delta t)$ decays at least an order of magnitude later than $p^{con}(\Delta t)$. This indicates that a lithium ion frequently jumps back into its initial layer before it finally leaves it. Probably, the reason is that the lithium ions are pulled back by the polymer. The slow decaying tails of $p^{dis}(\Delta t)$ at long lag times might also be caused by the finite size of the simulation box. Assuming statistical thermal motion of the lithium ions in the long term, a lithium ion will return to its initial layer after some time simply due to statistical reasons. The smaller the simulation box, i.e.\ the smaller the space a lithium ion can sample, the earlier the lithium ion will return.
\begin{table}[!ht]
  \centering
  \caption{Average continuous residence times $\tau_l^{con}$ of lithium ions in $z$-bins that can be attributed to actual \Lip{}-layers (see Figure~\ref{supp:fig:z-bins}) for the three investigated surface charges. The residence times are obtained from the continuous definition of the probability to stay in a given bin (see text for more information). All residence times are given in \si{\nano\second}.}
  \begin{tabular}{rrrr}
    & \multicolumn{3}{c}{$\sigma_S$ / \si{\elementarycharge\per\square\nano\meter}} \\
    Bin & $\pm 0$ & $\pm 0.5$ & $\pm 1$ \\ \hline
    $1$ &  $1.21$ &    $0.13$ &  $0.96$ \\
    $2$ &  $0.55$ &    $1.20$ &  $0.10$ \\
    $3$ &  $0.09$ &    $0.09$ &       - \\
    $4$ &  $0.08$ &         - &       - \\
    $5$ &  $0.10$ &         - &       - \\ \hline
    $14$ &       - &         - &  $0.11$ \\
    $15$ &  $0.07$ &         - &  $0.18$ \\
    $16$ &  $0.10$ &         - &  $0.52$ \\
    $17$ &  $0.07$ &    $0.60$ &  $4.94$ \\
    $18$ &  $0.61$ &   $11.06$ & $13.00$ \\
    $19$ &  $1.30$ &         - &       - \\
  \end{tabular}
  \label{supp:tab:lifetime_layer}
\end{table}

The continuous probabilities can be fitted well by stretched exponentials. The error in the obtained continuous residence times $\tau_l^{con}$ can be estimated very roughly from the inherent symmetry of the system with uncharged surfaces. A comparison of the residence times in equivalent layers (i.e.\ in Table~\ref{supp:tab:lifetime_layer} Bin~$1$ and Bin~$19$, Bin~$2$ and Bin~$18$ and so forth) indicates that the error in $\tau_l^{con}$ is on the order of $\pm \SI{10}{\percent}$.

The average continuous residence times of lithium ions in the different \Lip{}-layers increase with decreasing distance to the surface, regardless of the surface charge. This agrees well with the decreasing $z$-displacement variance (Figure~\ref{fig:displvar_cross_section} in the main paper) and can be explained by the increasing free energy barriers between the layers when approaching the surface (Figure~\ref{fig:free_energy_z} in the main paper). The average continuous residence time in \Lip{}-layers at uncharged and positive surfaces is similar (around \SI{1}{\nano\second}), again in agreement with the similar decrease of the $z$-displacement variance and the similar free energy barriers at these surfaces. At negative surfaces, the average continuous residence time is an order of magnitude larger, which conforms to the steeper decrease of the $z$-displacement variance and the higher free energy barriers at these surfaces.

\subsubsection*{On the Electrode Surface}

Similar to the layer residence times, the mean residence time of a lithium ion in the center of a graphene hexagon in the first layer at the negative electrode with $\sigma_S = -\SI{1}{\elementarycharge\per\square\nano\meter}$ can be computed (compare Figure~\ref{fig:densmap-xy_negative_electrode}a in the main paper). Here we are interested in the lithium-ion dynamics on the surface, i.e.\ for how long a lithium ion resides at a specific hexagonal site before it jumps to another hexagonal site. Jumps out of the first layer should not contribute. Therefore, if a lithium ion jumps out of the first layer, it does not contribute to the denominator $N_{L_i}(t_0)$ of Equations~\eqref{supp:eq:lifetime_continuous} and \eqref{supp:eq:lifetime_discontinuous}, although it was at a hexagonal site at time $t_0$ ($L_i$ stands now for the $i$-th hexagonal site instead of the $i$-th \Lip{}-layer).
\begin{figure}[!ht]
  \centering
  \includegraphics{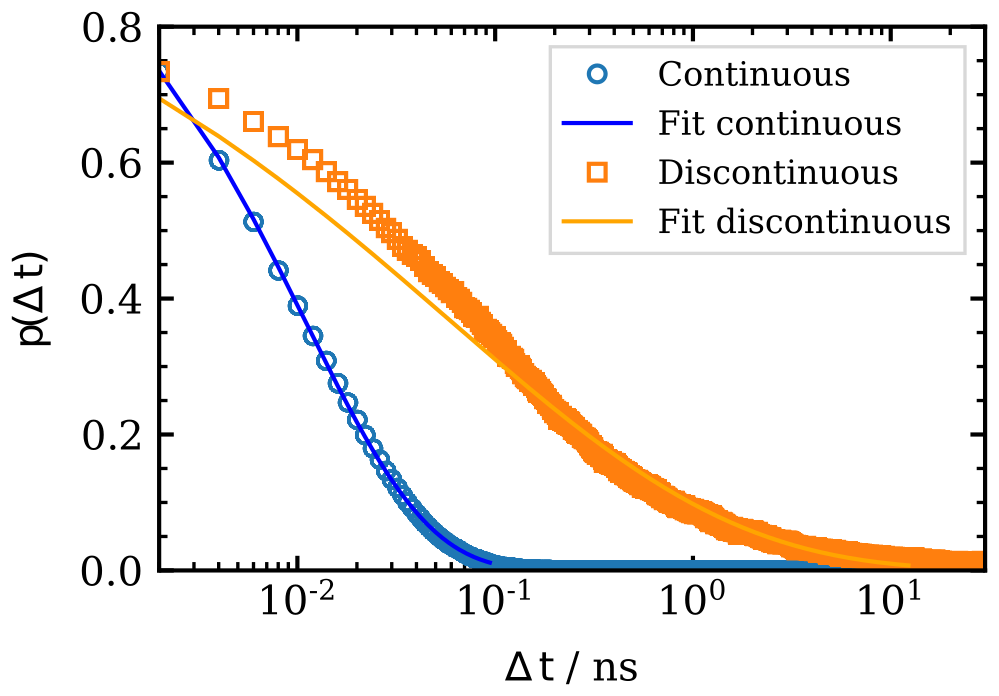}
  \caption{Probability that a lithium ion stays on a specific hexagonal site in the first layer at the negative electrode with $\sigma_S = -\SI{1}{\elementarycharge\per\square\nano\meter}$ (as shown in Figure~\ref{fig:densmap-xy_negative_electrode}a in the main paper) as function of lag time $\Delta t$. The blue points show the results for the continuous definition of the probability, the orange squares show the results for the discontinuous definition (see text for more information). The solid lines are stretched exponential fits to the data until the probability drops below $0.01$ or until $\SI{90}{\percent}$ of the maximum lag time of $\SI{1e3}{\nano\second}$ are reached (whatever happens earlier).}
  \label{supp:fig:lifetime_hex}
\end{figure}

The continuous and discontinuous probability for a lithium ion to stay on a given hexagonal site as function of lag time are shown in Figure~\ref{supp:fig:lifetime_hex}. The associated mean residence times are $\tau_h^{con} = \SI{0.01}{\nano\second}$ and $\tau_h^{dis} = \SI{0.56}{\nano\second}$, which are orders of magnitude shorter than the average continuous residence time in the first \Lip{}-layer ($\tau_l^{con} = \SI{13.00}{\nano\second}$, see Table~\ref{supp:tab:lifetime_layer}). This agrees qualitatively with the faster parallel \Lip{}-displacement variance compared to the perpendicular one (Figure~\ref{fig:displvar_cross_section} in the main paper) and with the smaller free energy barriers for lithium-ion diffusion on the negative surface compared to the barrier for leaving the first \Lip{}-layer (Figures~\ref{fig:free_energy_z} and \ref{fig:free_energy_hex} in the main paper). Note, however, that the hexagonal residence times can not be compared exactly with the layer residence time due to the different travel distance associated with these timescales. The distance between two hexagonal sites is given by the lattice constant of graphene, which is $r_0 \sqrt{3} = \SI{0.246}{\nano\meter}$. The distance between the maxima of the first and second \Lip{}-layers is $\SI{0.355}{\nano\meter}$.

The ratio of the hexagonal residence times $\tau_h^{dis}/\tau_h^{con}$ is $56$, indicating that a lithium ion jumps approximately $55$ times back to its initial hexagonal site before it finally leaves it. In the main paper we have shown that the lithium-ion dynamics on the negative surface is also governed by the polymer dynamics like in the bulk (Figure~\ref{fig:displvar_time_tscaled} in the main paper). Thus, the frequent back jumps might be explained by the polymer pulling back the lithium ions to their original site.

\subsection*{Role of Interchain Transfers for Layer Crossing}

\begin{figure}[H]
  \centering
  \includegraphics[width=\textwidth]{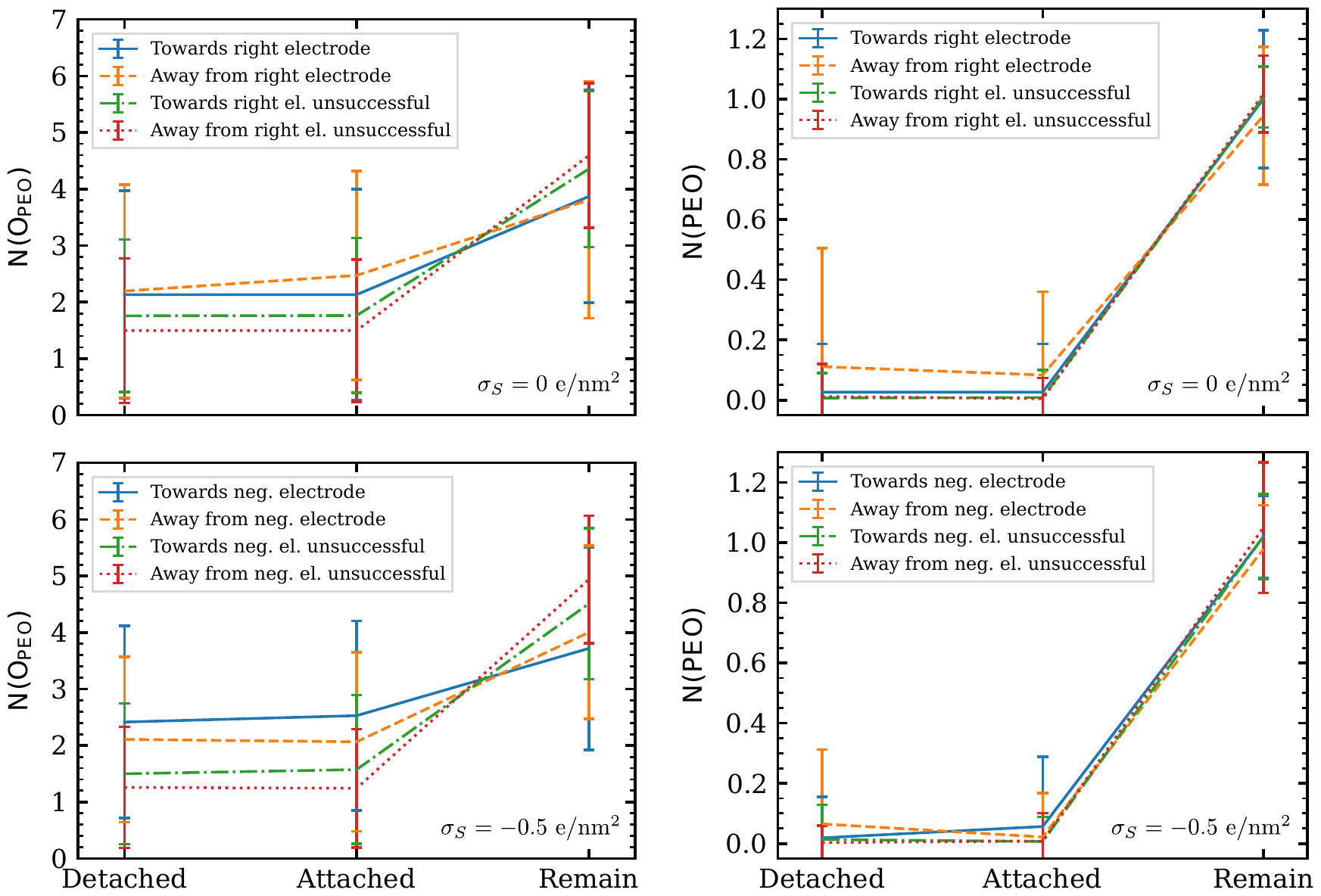}
  \caption{Upper panel: Exchange of ether oxygens from PEO (left) and of entire PEO chains (right) that coordinate to lithium ions that cross the free energy barrier between the first and second \Lip{}-layer ($z = \SI{9.5}{\nano\meter}$) at the right electrode in the system with $\sigma_S = \SI{0}{\elementarycharge\per\square\nano\meter}$. Lower panel: Same for lithium ions that cross the free energy barrier between the first and second \Lip{}-layer ($z = \SI{9.5}{\nano\meter}$) at the electrode with $\sigma_S = -\SI{0.5}{\elementarycharge\per\square\nano\meter}$. The error bars indicate the standard deviation of the coordination numbers. The corresponding plots for $\sigma_S = -\SI{1}{\elementarycharge\per\square\nano\meter}$ are shown in Figure~\ref{fig:lig_change_at_pos_change} in the main paper.}
  \label{supp:fig:lig_change_at_pos_change_near_negative_electrode}
\end{figure}

\begin{figure}[H]
  \centering
  \includegraphics[width=\textwidth]{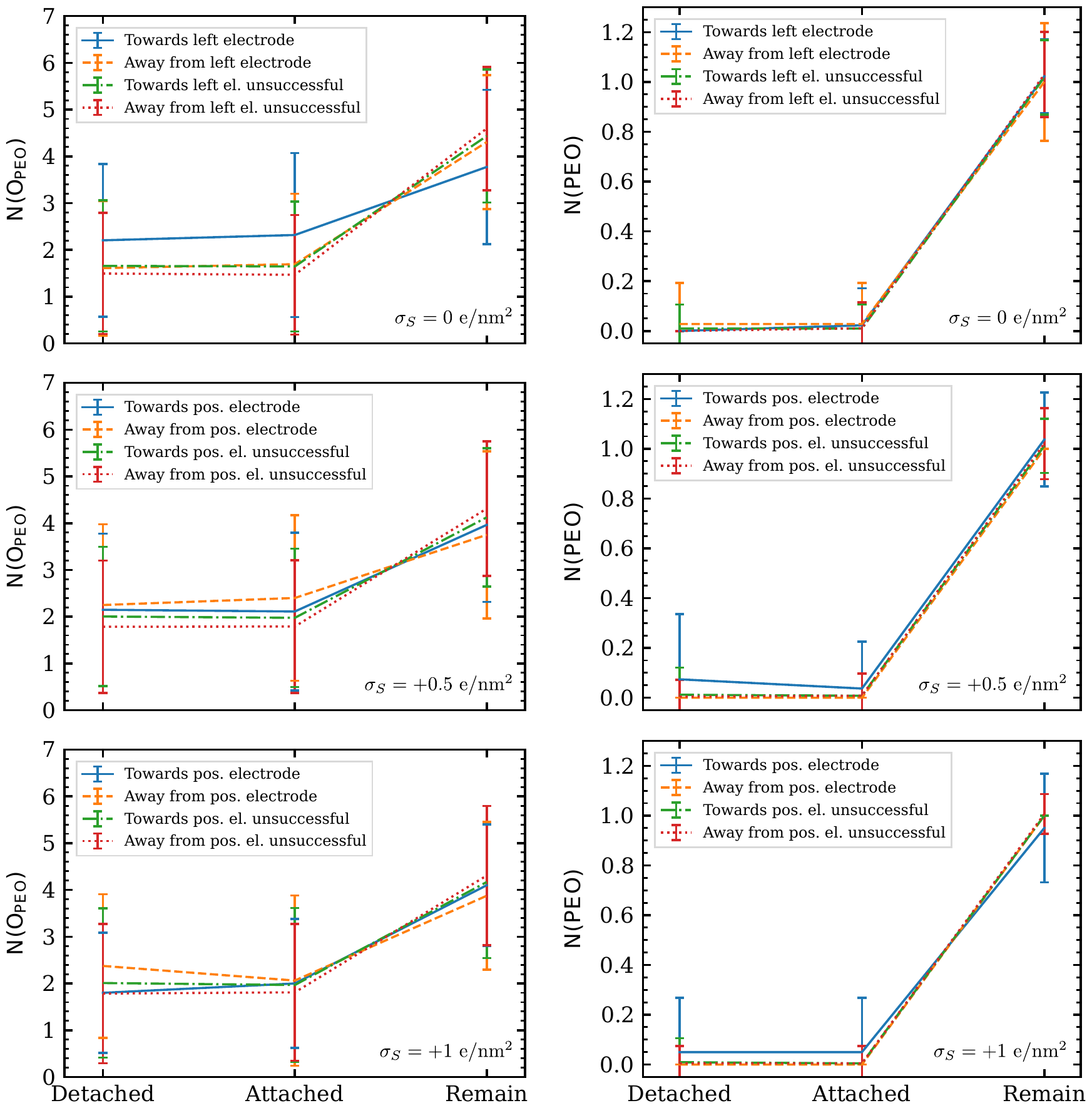}
  \caption{Top panel: Exchange of ether oxygens from PEO (left) and of entire PEO chains (right) that coordinate to lithium ions that cross the free energy barrier between the first and second \Lip{}-layer ($z = \SI{0.8}{\nano\meter}$) at the left electrode in the system with $\sigma_S = \SI{0}{\elementarycharge\per\square\nano\meter}$. Mid panel: Same for lithium ions that cross the free energy barrier between the first and second \Lip{}-layer ($z = \SI{1.3}{\nano\meter}$) at the electrode with $\sigma_S = +\SI{0.5}{\elementarycharge\per\square\nano\meter}$. Bottom panel: Same for lithium ions that cross the free energy barrier between the first and second \Lip{}-layer ($z = \SI{1.4}{\nano\meter}$) at the electrode with $\sigma_S = +\SI{1}{\elementarycharge\per\square\nano\meter}$. The error bars indicate the standard deviation of the coordination numbers.}
  \label{supp:fig:lig_change_at_pos_change_near_positive_electrode}
\end{figure}

\begin{table}[H]
  \centering
  \caption{Number of counted layer crossing events, $N_{cross}$, and average crossover time of the counted events, $t_{cross}$, for transitions between the first two layers at neutral, positive and negative electrodes. See Figure~\ref{fig:lig_change_at_pos_change_definitions} in the main paper for the definition of a counted  crossing event and the crossover time. The abbreviations \enquote{el.} and \enquote{unsucc.} stand for \enquote{electrode} and \enquote{unsuccessful}, respectively.}
  \begin{tabular*}{\textwidth}{l @{\extracolsep{\fill}} rrrr}
    & Towards el. & Away from el. & Towards el. unsucc. & Away from el. unsucc.                   \\ \hline
    & & & & \\
    & \multicolumn{4}{c}{Neutral and negative electrodes} \\ \hline
    & \multicolumn{4}{c}{$\sigma_S = \SI{0}{\elementarycharge\per\square\nano\meter}$, Barrier at $z = \SI{9.5}{\nano\meter}$}    \\
    $N_{cross}$                     &        $38$ &          $36$ &               $585$ &                                   $422$ \\
    $t_{cross}$ / \si{\pico\second} & $28 \pm 17$ &   $23 \pm 20$ &         $14 \pm 16$ &                             $18 \pm 18$ \\ \hline
    & \multicolumn{4}{c}{$\sigma_S = -\SI{0.5}{\elementarycharge\per\square\nano\meter}$, Barrier at $z = \SI{9.5}{\nano\meter}$} \\
    $N_{cross}$                     &        $53$ &          $46$ &               $148$ &                                   $933$ \\
    $t_{cross}$ / \si{\pico\second} & $17 \pm 17$ &   $34 \pm 23$ &         $13 \pm 16$ &                             $17 \pm 17$ \\ \hline
    & \multicolumn{4}{c}{$\sigma_S = -\SI{1}{\elementarycharge\per\square\nano\meter}$, Barrier at $z = \SI{9.9}{\nano\meter}$}   \\
    $N_{cross}$                     &       $446$ &         $456$ &               $215$ &                                  $1056$ \\
    $t_{cross}$ / \si{\pico\second} & $46 \pm 12$ &   $51 \pm 12$ &         $16 \pm 18$ &                             $13 \pm 15$ \\
    & & & & \\
    & \multicolumn{4}{c}{Neutral and positive electrodes} \\ \hline
    & \multicolumn{4}{c}{$\sigma_S = \SI{0}{\elementarycharge\per\square\nano\meter}$, Barrier at $z = \SI{0.8}{\nano\meter}$}    \\
    $N_{cross}$                     &        $44$ &          $36$ &               $525$ &                                   $366$ \\
    $t_{cross}$ / \si{\pico\second} & $32 \pm 21$ &   $27 \pm 20$ &         $17 \pm 17$ &                             $19 \pm 19$ \\ \hline
    & \multicolumn{4}{c}{$\sigma_S = +\SI{0.5}{\elementarycharge\per\square\nano\meter}$, Barrier at $z = \SI{1.3}{\nano\meter}$} \\
    $N_{cross}$                     &        $27$ &          $20$ &               $250$ &                                   $866$ \\
    $t_{cross}$ / \si{\pico\second} & $43 \pm 23$ &   $31 \pm 16$ &         $19 \pm 18$ &                             $19 \pm 19$ \\ \hline
    & \multicolumn{4}{c}{$\sigma_S = +\SI{1}{\elementarycharge\per\square\nano\meter}$, Barrier at $z = \SI{1.4}{\nano\meter}$}   \\
    $N_{cross}$                     &        $20$ &          $16$ &               $213$ &                                   $616$ \\
    $t_{cross}$ / \si{\pico\second} & $34 \pm 21$ &   $39 \pm 24$ &         $21 \pm 19$ &                             $19 \pm 18$ \\
  \end{tabular*}
  \label{supp:tab:lig_change_at_pos_change}
\end{table}
The total number of crossing events between the first two layers at the negative electrode with $\sigma_S = -\SI{1}{\elementarycharge\per\square\nano\meter}$ is much higher than at all other electrodes ($2173$ compared to around $1000$), because more crossing events fulfill the $\SI{1}{\nano\second}$ waiting time that a lithium ion has to stay continuously on one side of the barrier, due to the higher residence times of lithium ions in these two layers (Table~\ref{supp:tab:lifetime_layer}), and also because of the higher \Lip{} density at this electrode.

At uncharged surfaces, the lower number of unsuccessful jumps away from the electrode, compared to jumps towards the electrode, agrees with the lower energy barrier for jumps away from the electrode (Figure~\ref{fig:free_energy_z} in the main paper). At charged surfaces, the opposite is true.

\clearpage
\section*{Force Field Validation}

In a previous publication we have shown that MD simulations of \LiTFSI{}-glyme solvate ionic liquids with ion charges scaled by a factor of $0.8$ yield significantly better results with respect to experiments than simulations without charge scaling \cite{Thum2020}. In the study at hand we use the same force fields and charge scaling, but because the investigated PEO-\LiTFSI{} system is conceptually very different from the \LiTFSI{}-glyme system (the former being a polymer electrolyte and the latter a solvate ionic liquid), we validated the charge scaling approach again by comparing it to simulations using the APPLE\&P polarizable force field \cite{Borodin2006_APPLEP, Borodin2006_APPLEPa, Borodin2009_APPLEP} and to experiments. Regarding the APPLE\&P force field, we compared our simulations to simulations utilizing a newer  (\enquote{APPLE\&P new}) and an older (\enquote{APPLE\&P old}) version of the force field. These simulations were conducted by one of the authors in the context of other studies \cite{Diddens2013, Diddens2017}.

We also tested different charge scaling schemes to examine the influence of scaled charges on some key structural and dynamic properties. Namely, we performed simulations with the partial charges of all atoms scaled by a factor of $0.8$ (\enquote{Global $0.8$ $q$}), simulations with only the partial charges of atoms belonging to ionic species scaled by a factor of $0.8$ (\enquote{Ions $0.8$ $q$}) or $0.9$ (\enquote{Ions $0.9$ $q$}) and simulations with unscaled charges (\enquote{Unscaled $q$}).

Additionally, we compared two different sets of Lennard-Jones parameters for lithium ions. The first set ($\sigma = \SI{0.125992}{\nano\meter}$, $\epsilon = \SI{2.61500e+01}{\kilo\joule\per\mole}$) originates from Chandrasekhar \textit{et al.} \cite{Chandrasekhar1984_OPLS-AA} and is available in the OPLS-AA force field \cite{Jorgensen1996_OPLS-AA} as atom-type opls404. The second set ($\sigma = \SI{0.212645}{\nano\meter}$, $\epsilon = \SI{7.64793e-02}{\kilo\joule\per\mole}$) was introduced by Aqvist \cite{Aqvist1990_OPLS-AA} and has atom-type opls406 in the OPLS-AA force field. The lithium parameters by Aqvist are also used by the CL\&P force filed \cite{CanongiaLopes2012_CLP}.

The results shown in the main paper are derived from simulations with atomic point charges of ionic species scaled by a factor of $0.8$ and utilizing atom-type opls406. In the following tables, this parameter combination is highlighted in bold font. All results that are shown here are derived from bulk simulations. The simulation setup was the same as described in the main paper.

\subsection*{Densities}

Table~\ref{supp:tab:densities} lists the densities calculated for the different force field parameters from the last $\SI{10}{\nano\second}$ of the last equilibration step in an $NpT$ ensemble (see \enquote{Simulation details} section in the main paper). Clearly, the density decreases with increasing charge scaling, which can be explained by the resulting decrease of the electrostatic attraction between oppositely charged atoms. Moreover, the density of systems using opls406-lithium is less than the density of systems using opls404-lithium, which can be attributed to the larger Lennard-Jones radius~$\sigma$ of opls406-lithium. The density from the simulation utilizing opls404-lithium with ion charges scaled by a factor of $0.8$ is closest to the density from a simulation using the new APPLE\&P force field \cite{Diddens2017}. Generally, the simulations with scaled ion charges yield a better density approximation than the simulations with no or globally scaled charges, when taking the APPLE\&P polarizable force field as reference.
\begin{table}[!ht]
  \centering
  \caption{Densities of PEO-\LiTFSI{} electrolytes (EO:Li = 20:1) at $\SI{423}{\kelvin}$ calculated for different force field parameters from the last $\SI{10}{\nano\second}$ of the last equilibration step in an $NpT$ ensemble (see \enquote{Simulation details} section in the main paper). $\Delta\rho$ is the relative deviation to the density obtained from the new APPLE\&P force field. The parameter combination used for the main paper is highlighted in bold font.}
  \begin{tabular*}{\textwidth}{l @{\extracolsep{\fill}} rr @{\hskip 8\tabcolsep} rr}
    &                                      \multicolumn{2}{r@{\hskip 8\tabcolsep}}{\Lip{} opls404} &                              \multicolumn{2}{r}{\Lip{} opls406 / CL\&P} \\
    &               \multicolumn{2}{r@{\hskip 8\tabcolsep}}{$\sigma = \SI{0.125992}{\nano\meter}$} &               \multicolumn{2}{r}{$\sigma = \SI{0.212645}{\nano\meter}$} \\
    & \multicolumn{2}{r@{\hskip 8\tabcolsep}}{$\epsilon = \SI{2.61500e+01}{\kilo\joule\per\mole}$} & \multicolumn{2}{r}{$\epsilon = \SI{7.64793e-02}{\kilo\joule\per\mole}$} \\
    & & & & \\
    &                       $\rho$ / \si{\kilogram\per\cubic\meter} & $\Delta\rho$ / \si{\percent} &  $\rho$ / \si{\kilogram\per\cubic\meter} & $\Delta\rho$ / \si{\percent} \\ \hline
    Global $0.8$ $q$                &                                                     $1132.80$ &                      $-1.58$ &                                $1125.71$ &                      $-2.19$ \\
    Ions   $0.8$ $q$                &                                                     $1149.79$ &                      $-0.10$ &                           $\bm{1143.36}$ &                 $\bm{-0.66}$ \\
    Ions   $0.9$ $q$                &                                                     $1161.57$ &                       $0.92$ &                                $1153.59$ &                       $0.23$ \\
    Unscaled     $q$                &                                                     $1169.41$ &                       $1.60$ &                                $1164.24$ &                       $1.15$ \\
    & & & & \\
    APPLE\&P new \cite{Diddens2017} & \multicolumn{4}{c}{$\SI{1150.96}{\kilogram\per\cubic\meter}$} \\
  \end{tabular*}
  \label{supp:tab:densities}
\end{table}

\subsection*{Radial Distribution Functions}

\begin{figure}[!ht]
  \centering
  \includegraphics[height=0.909\textheight]{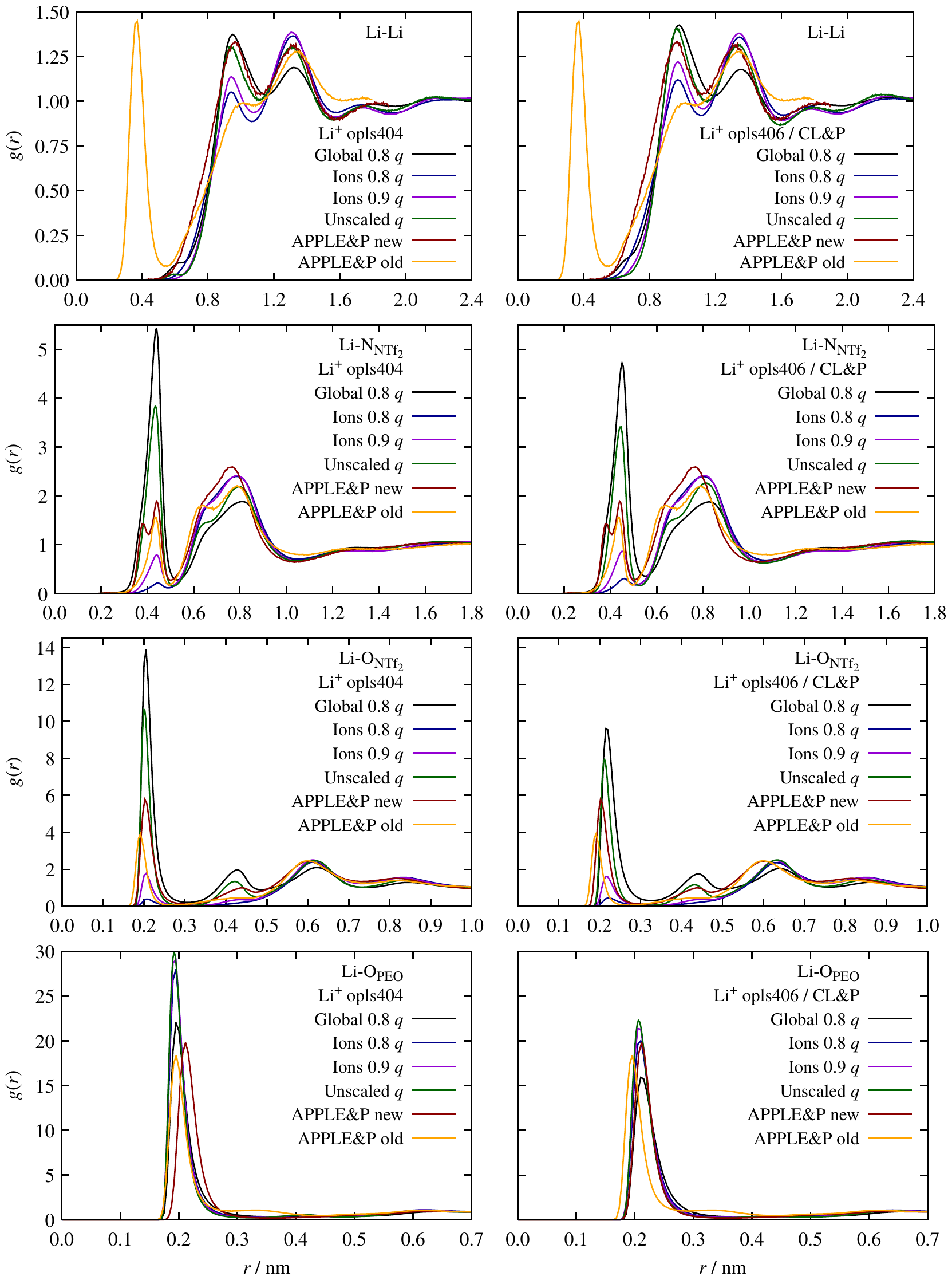}
  \caption{Radial distribution functions, $g(r)$, from simulations of PEO-\LiTFSI{} electrolytes (EO:Li = 20:1) with different force field parameters at $\SI{423}{\kelvin}$.}
  \label{supp:fig:rdf}
\end{figure}

Figure~\ref{supp:fig:rdf} compares radial distribution functions (RDFs) of selected atom pairs for different force field parameters. Overall, the positions of the peaks from the different simulations match well with each other, but the peak intensities partially show great discrepancies. Note, however, that the RDFs are normed by the density of the respective system. Hence, the peak intensities will naturally show at least small deviations. Instead of comparing the intensities of RDF peaks it is therefore better to compare coordination numbers, which will be done in the next section. Here we will focus only on the positions of the peaks.

As already said, the peak positions are overall in good agreement, though some deviations are present. The most obvious difference is the peak at around $\SI{0.4}{\nano\meter}$ in the Li-Li RDF of the old APPLE\&P force field. This peak seems rather spurious because it indicates the formation of clusters of lithium ions. Even when taking polarization into account, it is unlikely that lithium ions form clusters in this intermediately concentrated PEO-\LiTFSI{} polymer electrolyte with an ether oxygen to lithium ratio of EO:Li = 20:1. The fact that this peak is not present in simulations using a newer version of the APPLE\&P force field supports the hypothesis that this peak is an artifact.

Another obvious difference is the double peak structure in the Li-\NBT{} RDF of the new APPLE\&P force field. The first branch of this double peak indicates that a significant amount of \TFSI{} anions coordinate to the same lithium ion with two oxygen atoms \cite{Lesch2016, Li2015}. Taking a look at the coordination histograms in Figure~\ref{supp:fig:coordination_histogram_Li-NTf2} reveals that this is indeed the case for around $\SI{5}{\percent}$ of all \TFSI{} anions.

Simulations with opls404-lithium yield almost the same RDFs as simulations with opls406-lithium. The main difference is that the peaks are slightly shifted to shorter distances in case of opls404-lithium due to the smaller Lennard-Jones radius~$\sigma$. Moreover, in the case of opls404-lithium the first Li-\NBT{} and Li-O peaks are sharper (that means higher and thinner), probably because of the deeper potential well-depth~$\epsilon$.

\subsection*{Cation-Anion Coordination Numbers}

\begin{figure}[!ht]
  \centering
  \includegraphics[width=\textwidth]{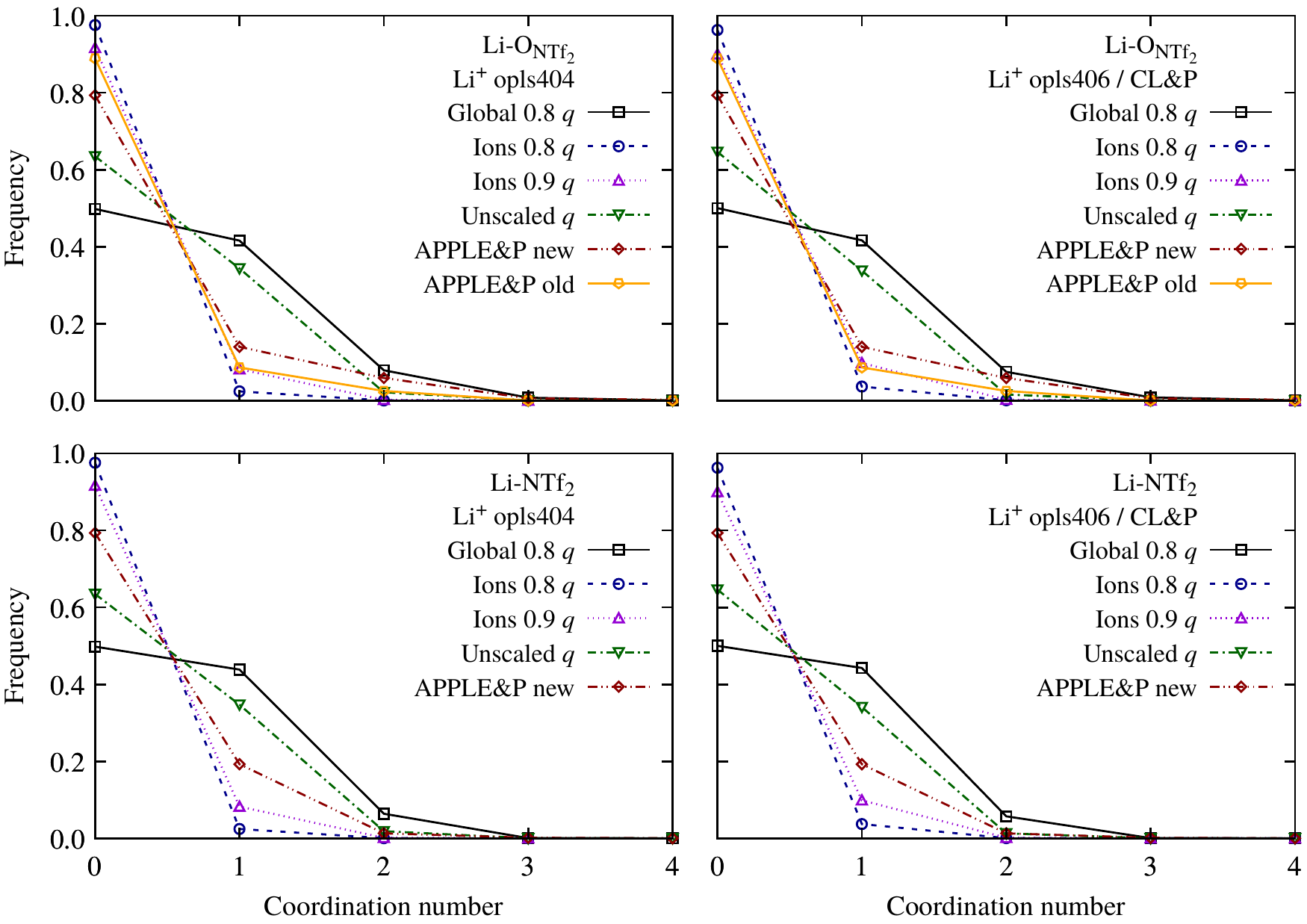}
  \caption{Coordination histograms for cation-anion coordination in simulations of PEO-\LiTFSI{} electrolytes (EO:Li = 20:1) with different force field parameters at $\SI{423}{\kelvin}$. The upper panel shows the frequency of specific coordination numbers for coordinations of lithium ions to single oxygen atoms from \TFSI{}. The lower panel shows the same but for coordinations to entire \TFSI{} anions.}
  \label{supp:fig:coordination_histogram_Li-NTf2}
\end{figure}

Figure~\ref{supp:fig:coordination_histogram_Li-NTf2} shows the frequency of specific cation-anion coordination numbers. A \TFSI{} anion is considered to coordinate to a lithium ion if at least one of its oxygen atoms is within a $\SI{0.3}{\nano\meter}$ cutoff radius of a lithium ion. This threshold corresponds to the first minimum in the Li-\OBT{} RDF (see Figure~\ref{supp:fig:rdf}).

In simulations with no or globally scaled charges, the cation-anion interaction is highly overestimated compared to the APPLE\&P polarizable force fields, meaning that there are significantly less free ions and more ion pairs. In the case of unscaled charges this can be attributed to the strong, non-reduced electrostatic attraction between cations and anions, which is reduced through polarization effects in the case of polarizable force fields. Surprisingly, the cation-anion interaction is also overestimated in the case of globally scaled charges, although the electrostatic attraction between cations and anions is reduced here. However, at the same time the electrostatic repulsion between like-charged ions is also reduced. This allows multiple counterions to coordinate to the same central-ion, as can for example be seen by the increased amount of lithium ions being coordinated by two \TFSI{} anions. Hence, in the case of globally scaled charges the formation of larger ion clusters is promoted.

Contrarily, reducing only the atomic point charges of ionic species does not promote the formation of ion clusters, although this leads to the same decrease of the electrostatic interactions between ions as in the case of globally scaled charges. However, the electrostatic interaction between the ions and the solvent is not decreased by the same amount, rendering ion-solvent interactions more favorable and cation-anion interactions less favorable. In fact, in simulations with reduced ion charges the cation-anion interaction is even slightly underestimated compared to simulations with the APPLE\&P polarizable force fields. This is especially visible in Table~\ref{supp:tab:coordination_numbers_Li-NTf2}, which lists the average cation-anion coordination numbers.
\begin{table}[!ht]
  \centering
  \caption{Average cation-anion coordination numbers in simulations of PEO-\LiTFSI{} electrolytes (EO:Li = 20:1) with different force field parameters at $\SI{423}{\kelvin}$. Li-\OBT{} denotes coordinations of lithium ions to single oxygen atoms from \TFSI{}. Li-\TFSI{} denotes coordinations of lithium ions to entire \TFSI{} anions. Additionally, the percentage of lithium-free \TFSI{} anions is given, which has also been estimated by Edman \cite{Edman2000} from Raman spectra of PEO-\LiTFSI{} electrolytes (EO:Li = 20:1) with $M_w = \SI{5e6}{\gram\per\mole}$ at $\SI{343}{\kelvin}$. The parameter combination used for the main paper is highlighted in bold font.}
  \begin{tabular*}{\textwidth}{l @{\extracolsep{\fill}} rr @{\hskip 6\tabcolsep} rr @{\hskip 6\tabcolsep} rr}
    & \multicolumn{2}{c@{\hskip 6\tabcolsep}}{Li-\OBT{}} & \multicolumn{2}{c@{\hskip 6\tabcolsep}}{Li-\TFSI{}} & \multicolumn{2}{c}{Li-free \TFSI{} / \si{\percent}} \\
    &                              opls404 &     opls406 &                              opls404 &     opls406  &                               opls404 &     opls406 \\ \hline
    Global $0.8$ $q$                             &                               $0.60$ &      $0.59$ &                                $0.57$ &      $0.56$ &                               $47.9$ &       $48.9$ \\
    Ions   $0.8$ $q$                             &                               $0.02$ & $\bm{0.04}$ &                                $0.02$ & $\bm{0.04}$ &                               $97.6$ &  $\bm{96.3}$ \\
    Ions   $0.9$ $q$                             &                               $0.09$ &      $0.10$ &                                $0.08$ &      $0.10$ &                               $91.6$ &       $90.0$ \\
    Unscaled     $q$                             &                               $0.39$ &      $0.37$ &                                $0.38$ &      $0.37$ &                               $64.0$ &       $65.4$ \\
    APPLE\&P new \cite{Diddens2017}              & \multicolumn{2}{c@{\hskip 6\tabcolsep}}{$0.29$}    & \multicolumn{2}{c@{\hskip 6\tabcolsep}}{$0.22$}     & \multicolumn{2}{c}{-}                               \\
    APPLE\&P old \cite{Diddens2013}              & \multicolumn{2}{c@{\hskip 6\tabcolsep}}{$0.14$}    & \multicolumn{2}{c@{\hskip 6\tabcolsep}}{-}          & \multicolumn{2}{c}{-}                               \\
    Raman ($\SI{343}{\kelvin}$) \cite{Edman2000} & \multicolumn{2}{c@{\hskip 6\tabcolsep}}{-}         & \multicolumn{2}{c@{\hskip 6\tabcolsep}}{-}          & \multicolumn{2}{c}{$96$}                            \\
  \end{tabular*}
  \label{supp:tab:coordination_numbers_Li-NTf2}
\end{table}

Additionally, Table~\ref{supp:tab:coordination_numbers_Li-NTf2} lists the fraction of \TFSI{} anions that are not coordinated to any lithium ion. From Raman spectra this fraction has been determined to be around $\SI{96}{\percent}$ in PEO-\LiTFSI{} electrolytes (EO:Li = 20:1) at $\SI{343}{\kelvin}$. With $\SI{96.3}{\percent}$ of lithium-free \TFSI{} anions, the simulation using opls406-lithium and ion charges scaled by a factor of $0.8$ resembles the experiment almost perfectly. However, one has to keep in mind the temperature difference of $\SI{80}{\kelvin}$ between experiment and simulation, making a direct comparison difficult.

Finally, Figure~\ref{supp:fig:coordination_histogram_Li-NTf2} and Table~\ref{supp:tab:coordination_numbers_Li-NTf2} show that there are almost no differences between opls404- and opls406-lithium regarding the cation-anion coordination.

\subsection*{Cation-Polymer Coordination Numbers}

\begin{figure}[!ht]
  \centering
  \includegraphics[width=\textwidth]{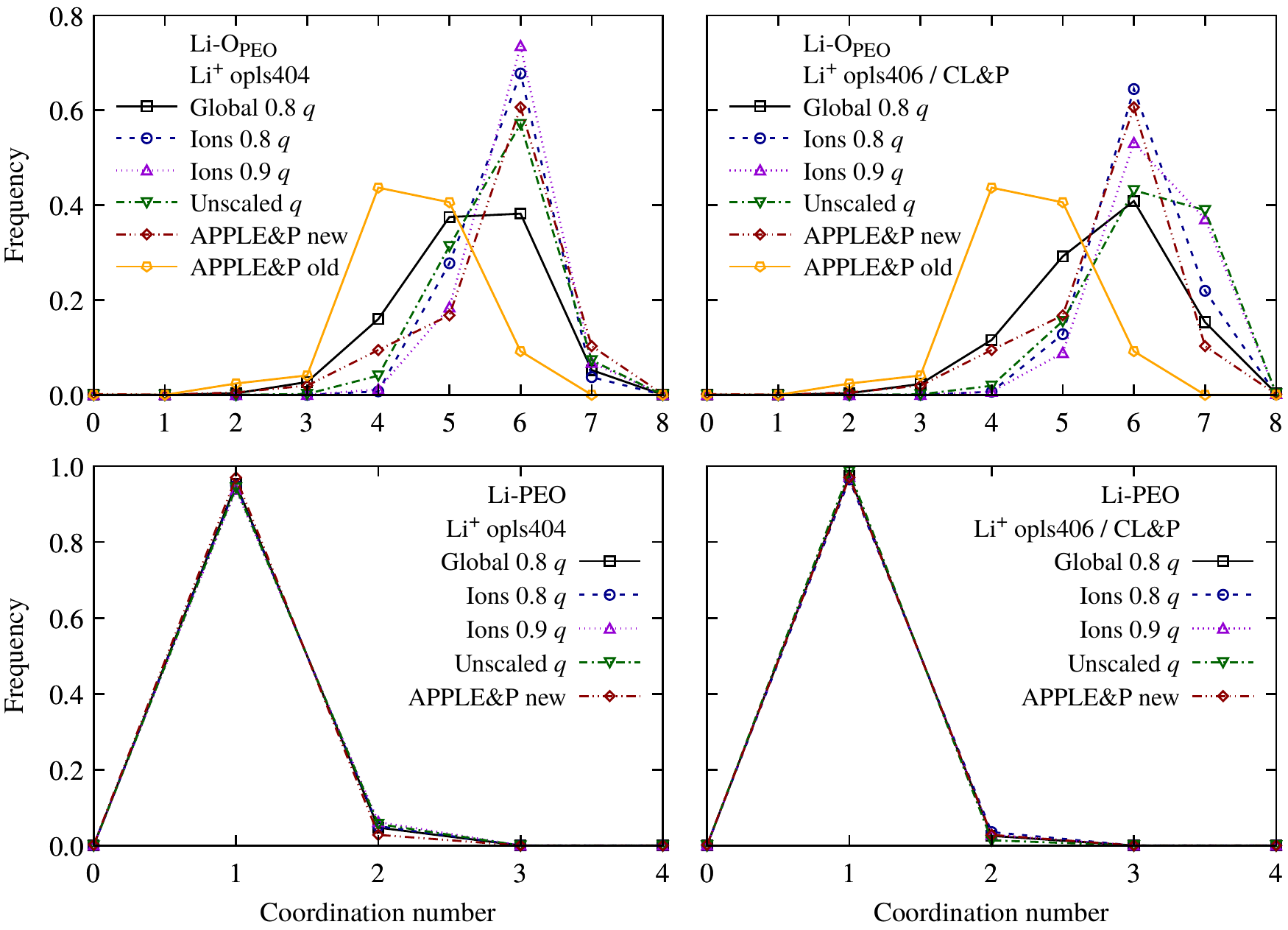}
  \caption{Coordination histograms for cation-polymer coordination in simulations of PEO-\LiTFSI{} electrolytes (EO:Li = 20:1) with different force field parameters at $\SI{423}{\kelvin}$. The upper panel shows the frequency of specific coordination numbers for coordinations of lithium ions to single oxygen atoms from PEO. The lower panel shows the same but for coordinations to entire PEO chains.}
  \label{supp:fig:coordination_histogram_Li-PEO}
\end{figure}

Figure~\ref{supp:fig:coordination_histogram_Li-PEO} shows the frequency of specific cation-polymer coordination numbers. A PEO chain is considered to coordinate to a lithium ion if at least one of its oxygen atoms is within a $\SI{0.3}{\nano\meter}$ cutoff radius of a lithium ion. This threshold corresponds to the first minimum in the Li-\OPEO{} RDF (see Figure~\ref{supp:fig:rdf}).

The lower panel of Figure~\ref{supp:fig:coordination_histogram_Li-PEO} shows that in all simulations almost all lithium ions are coordinated by one PEO chain. A few lithium ions are coordinated by two PEO chains, whereas there are no lithium ions that are not in contact with PEO. The Li-\OPEO{} coordination number distributions (upper panel of Figure~\ref{supp:fig:coordination_histogram_Li-PEO}), however, vary between the different simulations. As reflected in the average coordination numbers (Table~\ref{supp:tab:coordination_numbers_Li-PEO}), the distributions from the simulation with globally scaled charges and from the simulation using the old APPLE\&P force field are shifted to lower coordination numbers and do not have a very distinct maximum. All other distributions clearly peak at a Li-\OPEO{} coordination number of six.

The average coordination numbers in Table~\ref{supp:tab:coordination_numbers_Li-PEO} show that in all simulations a lithium ion is on average coordinated by five to six oxygen atoms from one PEO chain. The only exception is the simulation using the old APPLE\&P force field, which yields an average Li-\OPEO{} coordination number of $4.50$. Mao \textit{et al.} determined an average Li-\OPEO{} coordination number of $4.9 \pm 0.5$ using neutron diffraction with isotopic substitution (NDIS), which is lower than the average in all simulations with non-polarizable force fields. However, the experiment was done at $\SI{296}{\kelvin}$ and, more importantly, the lithium salt concentration was more than double as high, namely EO:Li = 7.5:1. This could explain the somewhat lower Li-\OPEO{} coordination number in the experiment, because the PEO chains are nearly saturated with lithium ions and the chance for a lithium ion to find an anion oxygen rather than an polymer oxygen is increased.
\begin{table}[!ht]
  \centering
  \caption{Average cation-polymer coordination numbers in simulations of PEO-\LiTFSI{} electrolytes (EO:Li = 20:1) with different force field parameters at $\SI{423}{\kelvin}$. Li-\OPEO{} denotes coordinations of lithium ions to single oxygen atoms from PEO. Li-PEO denotes coordinations of lithium ions to entire PEO chains. Additionally, the Li-\OPEO{} coordination number obtained by Mao \textit{et al.} \cite{Mao2000} from NDIS (Neutron Diffraction with Isotopic Substitution) experiments on PEO-\LiTFSI{} electrolytes (EO:Li = 7.5:1) with $M_w = \SI{52e3}{\gram\per\mole}$ at $\SI{296}{\kelvin}$ is given. The parameter combination used for the main paper is highlighted in bold font.}
  \begin{tabular}{l rr @{\hskip 6\tabcolsep} rr}
    & \multicolumn{2}{c@{\hskip 6\tabcolsep}}{Li-\OPEO{}}    & \multicolumn{2}{c}{Li-PEO} \\
    &                                  opls404 &     opls406 &      opls404 &     opls406 \\ \hline
    Global $0.8$ $q$                                       &                                   $5.26$ &      $5.55$ &       $1.05$ &      $1.03$ \\
    Ions   $0.8$ $q$                                       &                                   $5.75$ & $\bm{6.08}$ &       $1.05$ & $\bm{1.04}$ \\
    Ions   $0.9$ $q$                                       &                                   $5.86$ &      $6.27$ &       $1.06$ &      $1.03$ \\
    Unscaled     $q$                                       &                                   $5.67$ &      $6.20$ &       $1.06$ &      $1.01$ \\
    APPLE\&P new \cite{Diddens2017}                        & \multicolumn{2}{c@{\hskip 6\tabcolsep}}{$5.66$}        & \multicolumn{2}{c}{$1.03$} \\
    APPLE\&P old \cite{Diddens2013}                        & \multicolumn{2}{c@{\hskip 6\tabcolsep}}{$4.50$}        & \multicolumn{2}{c}{-}      \\
    NDIS ($\SI{296}{\kelvin}$, EO:Li=7.5:1) \cite{Mao2000} & \multicolumn{2}{c@{\hskip 6\tabcolsep}}{$4.9 \pm 0.5$} & \multicolumn{2}{c}{-}      \\
  \end{tabular}
  \label{supp:tab:coordination_numbers_Li-PEO}
\end{table}
%\begin{table}[!ht]
%  \centering
%  \caption{Percentage of PEO-free cations and cation-free PEO chains.}
%  \begin{tabular}{l@{\hskip 6\tabcolsep}rr@{\hskip 8\tabcolsep}rr}
%                     & \multicolumn{2}{c@{\hskip 8\tabcolsep}}{PEO-free \Lip{} / \si{\percent}} & \multicolumn{2}{c}{Li-free PEO / \si{\percent}} \\
%                     &                                                     opls404 &    opls406 &                            opls404 &    opls406 \\ \hline
%    Global $0.8$ $q$ &                                                       $0.0$ &      $0.0$ &                              $0.8$ &      $0.4$ \\
%    Ions   $0.8$ $q$ &                                                       $0.0$ & $\bm{0.0}$ &                              $0.1$ & $\bm{0.5}$ \\
%    Ions   $0.9$ $q$ &                                                       $0.0$ &      $0.0$ &                              $0.4$ &      $0.3$ \\
%    Unscaled     $q$ &                                                       $0.0$ &      $0.0$ &                              $0.0$ &      $1.3$ \\
%    APPLE\&P new     &                           \multicolumn{2}{c@{\hskip 8\tabcolsep}}{$0.2$} &                           \multicolumn{2}{c}{-} \\
%    APPLE\&P old     &                           \multicolumn{2}{c@{\hskip 8\tabcolsep}}{$0.0$} &                           \multicolumn{2}{c}{-} \\
%  \end{tabular}
%  \label{supp:tab:solvation}
%\end{table}

Contrarily to the cation-anion coordination, the choice between opls404- or opls406-lithium leads to a precise difference in the cation-polymer coordination. In simulations using opls406-lithium, the Li-\OPEO{} coordination numbers are slightly, but noticeably, higher. The Li-PEO coordination numbers are almost the same in all simulations.

\subsection*{Mean-Square Displacements and Diffusion Coefficients}

\begin{figure}[!ht]
  \centering
  \includegraphics[height=0.887\textheight]{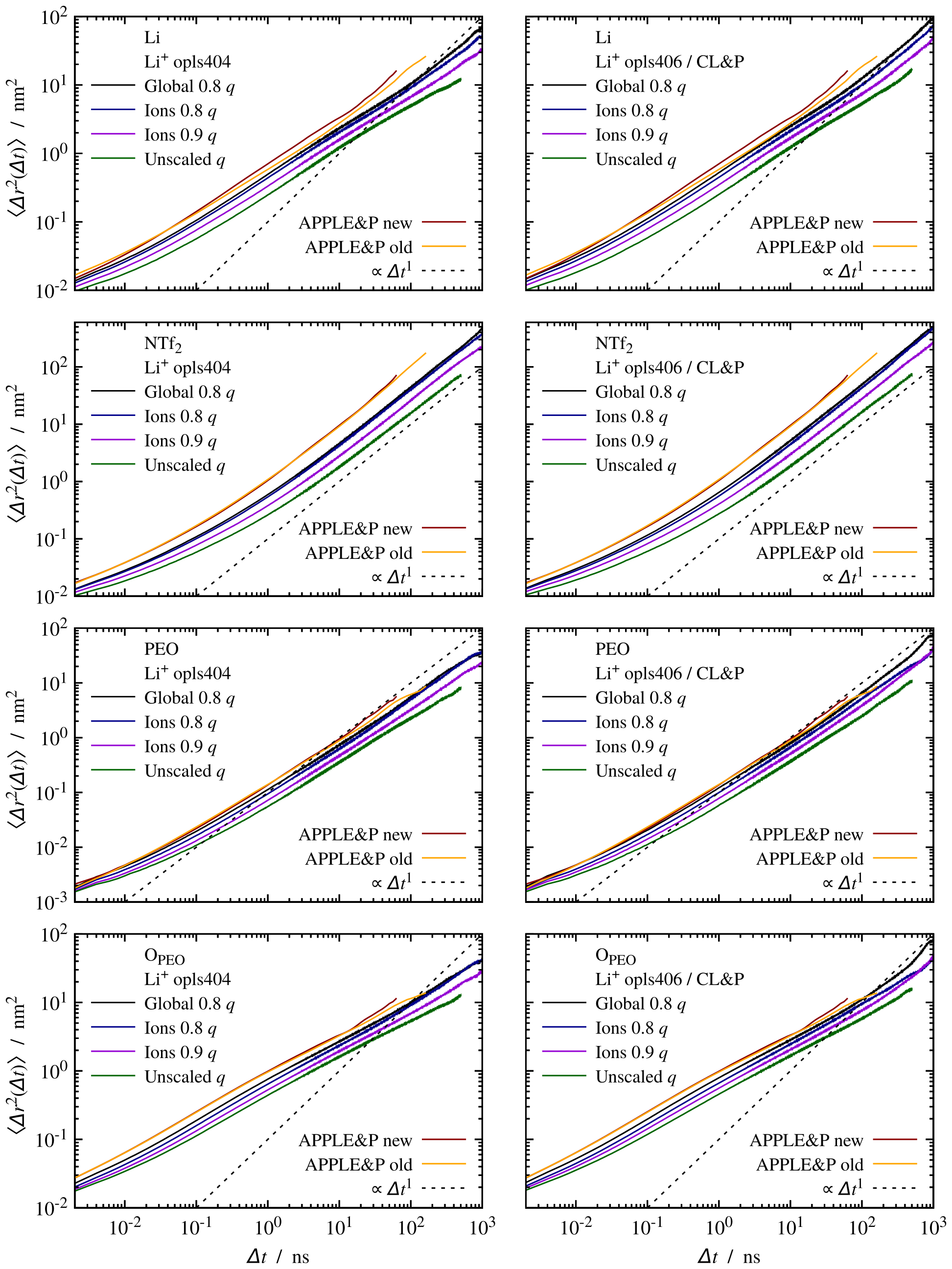}
  \caption{Mean-square displacements, $\langle \Delta \bm{r}^2(\Delta t) \rangle$, from simulations of PEO-\LiTFSI{} electrolytes (EO:Li = 20:1) with different force field parameters at $\SI{423}{\kelvin}$. The dashed lines indicate the diffusive regime.}
  \label{supp:fig:msd}
\end{figure}

Figure~\ref{supp:fig:msd} shows the mean-square displacements (MSDs) of lithium cations, the center of mass of \TFSI{} anions, the center of mass of PEO chains and the MSD of the ether oxygens of the PEO chains for different force field parameters. Opposing the trend of the densities, the MSDs of all species increase with increasing charge scaling, which is a direct consequence of the reduced electrostatic interaction. In all cases, the MSDs from the simulations with the APPLE\&P polarizable force fields are higher than the MSDs from the simulations with unpolarizable force fields. However, the difference to simulations with globally scaled charges or ion charges scaled by a factor of $0.8$ is small.

For \TFSI{} anions, simulations with globally scaled charges and simulations with ion charges scaled by a factor of $0.8$ yield almost identical MSDs. This can be related to the fact that \TFSI{} anions coordinate only to lithium cations but not to the polymer matrix. Thus, \TFSI{} anions only benefit from a reduction of ionic charges and are not further accelerated when additionally reducing the atomic point charges of the polymer.

A more quantitative picture of the influence of scaled charges on the dynamics of a system can be gained when looking at the self-diffusion coefficients in Table~\ref{supp:tab:diffusion}. The self-diffusion coefficient $D_\alpha$ of species $\alpha$ was obtained by fitting a straight line to the diffusive regime of the corresponding MSD according to the Einstein relation \cite{Frenkel2002}.
\begin{equation}
\lim_{\Delta t \to \infty} \langle \Delta \bm{r}_{\alpha,i}^2(\Delta t) \rangle = 6 D_\alpha \Delta t
\label{supp:eq:einstein_msd}
\end{equation}
Here, $\Delta \bm{r}_{\alpha,i}^2(\Delta t)$ denotes the MSD of the $i$-th molecule of species $\alpha$ after a diffusion time of $\Delta t$. It is defined as \cite{Frenkel2002}
\begin{equation}
\Delta \bm{r}_{\alpha,i}^2(\Delta t) = |\bm{r}_{\alpha,i}(t_0 + \Delta t) - \bm{r}_{\alpha,i}(t_0)|^2.
\label{supp:eq:msd}
\end{equation}
The brackets $\langle ... \rangle$ denote averaging over all molecules $i$ of species $\alpha$ and over multiple starting times $t_0$.
\begin{table}[!ht]
  \centering
  \caption{Self-diffusion coefficients from simulations of PEO-\LiTFSI{} electrolytes (EO:Li = 20:1) with different force field parameters at $\SI{423}{\kelvin}$. Additionally, the self-diffusion coefficients of cations and anions obtained by Gorecki \textit{et al.} \cite{Gorecki1995} from PFG-NMR (pulsed-field gradient nuclear magnetic resonance) experiments on PEO-\LiTFSI{} electrolytes (EO:Li = 20:1) with $M_w = \SI{9e5}{\gram\per\mole}$ at $\SI{400}{\kelvin}$ are given. The parameter combination used for the main paper is highlighted in bold font.}
  \begin{tabular*}{\textwidth}{l @{\extracolsep{\fill}} rr @{\hskip 4\tabcolsep} rr @{\hskip 4\tabcolsep} rr}
    & \multicolumn{2}{c@{\hskip 4\tabcolsep}}{$D($\Lip{}$)$ / \si{\square\angstrom\per\nano\second}} & \multicolumn{2}{c@{\hskip 4\tabcolsep}}{$D($\TFSI{}$)$ / \si{\square\angstrom\per\nano\second}} & \multicolumn{2}{c}{$D($PEO$)$ / \si{\square\angstrom\per\nano\second}} \\
    &                           opls404 &     opls406 &                            opls404 &     opls406 &      opls404 &     opls406 \\ \hline
    Global $0.8$ $q$                                 &                            $1.23$ &      $1.60$ &                             $7.37$ &      $8.29$ &       $0.88$ &      $1.08$ \\
    Ions   $0.8$ $q$                                 &                            $0.93$ & $\bm{1.20}$ &                             $6.55$ & $\bm{7.21}$ &       $0.75$ & $\bm{0.76}$ \\
    Ions   $0.9$ $q$                                 &                            $0.56$ &      $0.79$ &                             $4.35$ &      $4.61$ &       $0.46$ &      $0.63$ \\
    Unscaled     $q$                                 &                                 - &      $0.56$ &                             $2.51$ &      $2.72$ &       $0.27$ &      $0.37$ \\
    APPLE\&P new \cite{Diddens2017}                  & \multicolumn{2}{c@{\hskip 4\tabcolsep}}{$4.20$} & \multicolumn{2}{c@{\hskip 4\tabcolsep}}{$15.91$} & \multicolumn{2}{c}{$1.53$} \\
    APPLE\&P old \cite{Diddens2013}                  & \multicolumn{2}{c@{\hskip 4\tabcolsep}}{$3.20$} & \multicolumn{2}{c@{\hskip 4\tabcolsep}}{$15.48$} & \multicolumn{2}{c}{$1.25$} \\
    PFG-NMR ($\SI{400}{\kelvin}$) \cite{Gorecki1995} & \multicolumn{2}{c@{\hskip 4\tabcolsep}}{$6.03$} & \multicolumn{2}{c@{\hskip 4\tabcolsep}}{$14.86$} & \multicolumn{2}{c}{-}      \\
  \end{tabular*}
  \label{supp:tab:diffusion}
\end{table}

Table~\ref{supp:tab:diffusion} underlines again that the dynamics get faster with increasing charge scaling, i.e.\ with decreasing electrostatic interaction. For instance, the self-diffusion coefficient of lithium ions in simulations with unscaled charges is $\SI{0.56}{\square\angstrom\per\nano\second}$, more than an order of magnitude smaller than the experimental value of $\SI{6.03}{\square\angstrom\per\nano\second}$. When scaling the charges of ionic species by a factor of $0.8$, $D($\Lip{}$)$ increases to $\SI{1.20}{\square\angstrom\per\nano\second}$ and is now \enquote{only} five times smaller than the experimental value. Hence, charge scaling leads to self-diffusion coefficients with at least the correct order of magnitude.

From Table~\ref{supp:tab:diffusion} it becomes also obvious, that the dynamics are slightly faster in simulations using opls406-lithium compared to simulations using opls404-lithium. This is probably caused by the less steep potential well-depth~$\epsilon$ of opls406-lithium and its larger Lennard-Jones radius~$\sigma$, which gives rise to lower densities.

\subsection*{Conclusions from the Force Field Comparison}

Compared to the newer version of the APPLE\&P polarizable force field, simulations with only the charges of ionic species scaled by a factor of $0.9$ or $0.8$ yield the best results concerning the structural properties of the PEO-\LiTFSI{} electrolyte under consideration. Charge scaling by a factor of $0.9$ results in a better description of the cation-anion coordination, whereas charge scaling by a factor of $0.8$ better describes the cation-polymer coordination. Concerning the dynamics, global charge scaling yields the best results, although the differences to simulations with ion charges scaled by a factor of $0.8$ are small. The difference between opls404-lithium and opls406-lithium is small compared to the effects of charge scaling, although a systematic impact on structural and dynmaic properties was found.

We finally decided to use a charge scaling factor of $0.8$ for ionic species in combination with opls406-lithium for the following four reasons. i)~From the discussion above it appears that this parameter combination is the best compromise between structural and dynamic properties. ii)~Recently, we have used the same parameter combination to describe \LiTFSI{}-glyme solvate ionic liquids \cite{Thum2020}. It seems that this parameter combination can be used to model a variety of PEO-\LiTFSI{} electrolytes with a broad range of chain lengths and concentrations with a fair accuracy. iii)~The parameters of opls406-lithium are also used in the CL\&P force field \cite{CanongiaLopes2012_CLP}. Because we took the parameters for \TFSIm{} anion from this force field, it is more consistent to take the parameters for the cation from the same force field as well. iv)~In their theory about how to account for electronic polarization in non-polarizable force fields, Leontyev and Stuchebrukhov argue that the atomic point charges of ionic species should be scaled by a system-dependent factor, which is typically around $0.7$ \cite{Leontyev2011}. In our case we chose $0.8$.

Finally, we would like to remark that the APPLE\&P force field is also only a classical force field like the OPLS-AA force field. Although it is expected to give more accurate results due to the explicit incorporation of polarization, it still relies on model assumptions and is only as good as its parameterization. A thorough examination of the goodness of a force field can only be made in comparison to experimental data. Unfortunately, experimental data are not always available at the same conditions as used in simulations. In simulations, it is usually necessary to apply a high temperature to accelerate the slow polymer dynamics and hence to sample a sufficient portion of the phase space. In experiments, these high temperatures are not always accessible, because the polymer might decompose or the technical effort would be too high.

\putbib[literature_supplementary_bibtex]
\end{bibunit}

\end{document}